\documentclass[11pt,a4paper,onecolumn]{article}

\usepackage[left=1.6cm,right=1.6cm,top=1.6cm,bottom=2cm]{geometry}
\usepackage{natbib}
\usepackage{float}

\usepackage{amsmath,amssymb,amsfonts,amsthm}
\usepackage{graphicx}
\usepackage{xcolor}
\newtheorem{assumption}{Assumption}
\newtheorem{lemma}{Lemma}
\newtheorem{theorem}{Theorem}
\newtheorem{remark}{Remark}
\newtheorem{proposition}{Proposition}

\DeclareMathOperator{\rank}{rank}

\newcommand{\R}{\mathbb{R}}
\newcommand{\Sg}{\mathcal{S}_{g}}
\newcommand{\KKT}{\mathcal{X}_{\mathrm{KKT}}}
\newcommand{\vpd}{\mathbf v_{\mathrm{nom}}}
\title{\huge Safe Output-Regulation Flows for Constrained Optimization via Control Barrier Functions}
\author{%
    Zihao Ren,
    Bingqi Liu,
    Lei Wang, Hongye Su and
    Guodong Shi\thanks{Zihao Ren, Lei Wang and Hongye Su are with the College of Control Science and Engineering, Zhejiang University, P.R. China. (E-mail: zhren2000; lei.wangzju; hysu69@zju.edu.cn). Bingqi Liu and Guodong Shi are with the Australian Centre for Field Robotics, The University of Sydney, Australia. (Email: bliu0073; guodong.shi@sydney.edu.au).  
}}

\date{}

\begin{document}
\maketitle

\begin{abstract}
Control-theoretic approaches to constrained optimization have been widely studied by interpreting optimization algorithms as dynamical systems. This paper formulates constrained optimization as an output-regulation problem with state constraints, where equality-constraint residuals are treated as regulated outputs and inequality constraints as state constraints. Proportional--integral feedback is employed to drive the equality-constraint residuals asymptotically to zero, while a control barrier function (CBF) safety filter is used to ensure inequality safety throughout the optimization process. We establish feasibility of the CBF filtering program throughout the inequality-feasible set and local exponential convergence of the resulting flow. Under stronger global conditions, we further prove convergence from an entire connected component of the inequality-feasible set. Since exact implementation of the flow requires solving a CBF quadratic program at each state, we further develop a flow that approximates the CBF solution through multiplier updates. Moreover, we provide a unified formulation encompassing the two proposed flows and projected primal--dual dynamics. Numerical examples demonstrate convergence, safety, and computational tradeoffs of the proposed flows.
\end{abstract}


\section{Introduction}

Constrained optimization concerns the minimization of an objective function subject to a set of constraints. It plays a fundamental role in many engineering applications, including real-time decision-making, control of dynamical systems, and signal processing
\cite{Feijer2010Stability,Cherukuri2016Asymptotic,Qu2019Exponential,Bertsekas1999Nonlinear}.
In these applications, constraints capture physical limitations, safety requirements, and operational specifications, ensuring that the resulting solutions are feasible and practically implementable
\cite{Ames2017Control,Wabersich2023Predictive,Rawlings2017Model,Nagurney1996Projected}. 
These practical requirements have motivated extensive research on constrained optimization methods
\cite{Arrow1958Gradient,Tang2020SemiGlobal}.

For constrained optimization algorithms, a fundamental objective is to drive the optimization variables toward optimal solutions, typically characterized by Karush--Kuhn--Tucker (KKT) conditions. Among the approaches for constrained optimization, output--regulation flows form a well-established class for achieving such convergence \cite{Niederlander2016Distributed,Zhang2018Projected,Cherukuri2017Saddle,Nedic2018Network}. In their basic form, these methods evolve the primal variables along a descent direction of the Lagrangian while dynamically updating the associated dual multipliers. For equality constraints, this mechanism admits an output-regulation interpretation, where the constraint residual $\mathbf h(\mathbf x)$ is regarded as an output to be regulated to zero, with the augmentation term providing proportional correction and an integral controller state supplying integral action. For inequality constraints, projected dual dynamics maintain nonnegative multipliers, whose values determine the influence of the corresponding constraints in the primal update. Under suitable assumptions, the stability and convergence of these methods to KKT equilibria have been established.

Although output--regulation flows provide convergence guarantees toward KKT equilibria, convergence alone is not sufficient for many practical applications \cite{Wabersich2021Learning,Ma2021SafeMPC,Prajna2007Safety}. In safety-critical or physically constrained systems, the optimization trajectory itself is often required to satisfy certain constraints during the transient process. In particular, inequality constraints typically represent hard physical limitations, safety requirements, or operational boundaries, such as collision avoidance constraints in multi-agent systems and state--input constraints in safety-critical control systems \cite{Rawlings2017Model,Wang2017Safety,Xiao2020HighOrder,Blackmore2011Autonomous}. Temporary violations of such constraints may lead to unsafe behaviors or infeasible operations, and therefore the inequality-feasible set should be preserved throughout the optimization trajectory.

To address transient feasibility, \cite{allibhoy2024control} develops a safe gradient flow by interpreting constrained optimization as a safe control problem. Their approach is based on control barrier functions (CBFs), which provide a control-theoretic mechanism for rendering a prescribed feasible set forward invariant through pointwise constraints on the system vector field \cite{Aubin1991Viability,Ames2019Theory}. In this formulation, the constraints define the safe set, while the negative gradient of the objective serves as the nominal optimization direction. The proposed dynamics are shown to preserve the forward invariance of the safe set, meaning that any trajectory initialized from a physically feasible state remains feasible throughout the evolution. This safe optimization perspective has been further extended to distributed network optimization and variational inequality problems through safe monotone flows \cite{Mestres2023Distributed,Allibhoy2025Anytime}. 

The safe gradient flow in \cite{allibhoy2024control} renders the
full feasible set forward invariant by incorporating both equality
and inequality constraints into the safety filter. Here we consider
applications in which equality constraints are steady-state
requirements \cite{OlfatiSaber2007Consensus,Nedic2010Distributed,Boyd2011Distributed,Feijer2010Stability,Shi2015Distributed}, while inequality constraints must hold throughout
the transient. This distinction motivates adopting a control-theoretic perspective that treats equality-constraint residuals as regulated outputs and inequality constraints as safety requirements on the state.

In this paper, we propose an inequality-CBF-filtered output--regulation (ICBF-OR) flow that regulates equality-constraint residuals through proportional--integral output feedback and enforces inequality constraints through a CBF safety filter. The resulting flow guarantees feasibility of the filtering program throughout the inequality-feasible set and renders this set forward invariant. We establish local exponential convergence and, under additional conditions, convergence from an entire connected component of the inequality-feasible set. We also develop an A-ICBF-OR flow that approximates the CBF quadratic-program solution through multiplier updates, and characterize its convergence and transient inequality-constraint violations. Finally, we provide a unified parameterized formulation encompassing ICBF-OR, A-ICBF-OR, and projected primal--dual dynamics and revealing their inherent connections.

The main contributions of this paper are threefold.
\begin{itemize}
    \item We propose the ICBF-OR flow, which combines proportional--integral output regulation of the equality residuals with an inequality-only CBF safety filter. The proposed flow ensures feasibility of the filtering program
throughout the inequality-feasible set and uses fewer constraints
in the CBF quadratic program than formulations that also include
equality conditions in the filter.

    \item We develop an approximate implementation of ICBF-OR, termed the A-ICBF-OR flow, based on continuous-time multiplier updates. It avoids solving the filtering quadratic program to completion at each state. We establish an $O(\tau)$ bound on transient inequality-constraint violations, where $\tau>0$ is the approximation parameter governing the time scale of the multiplier updates.

    \item We establish local exponential convergence of both ICBF-OR and A-ICBF-OR flows without requiring a combined LICQ on the equality and active inequality gradients. Under additional global conditions, we further prove convergence of the ICBF-OR flow from an entire connected component in the inequality-feasible set. 
\end{itemize}

The remainder of this paper is organized as follows. Section~\ref{sec:problem} presents the problem formulation and preliminaries. Section~\ref{sec:algorithm} introduces the ICBF-OR flow. Section~\ref{sec:convergence} establishes basic properties of the flow and develops the local and safe-component convergence results. Section~\ref{sec:cbf_qp_solution} develops the A-ICBF-OR flow and establishes its local convergence and constraint-violation bound. Section~\ref{subsec:relations_solution_methods} clarifies the relations among ICBF-OR, A-ICBF-OR, and projected primal--dual flows. Section~\ref{sec:numerical_simulations} presents numerical simulations, and Section~\ref{sec:conclusion} concludes the paper. Proofs are provided in the appendices.

\medskip

\noindent\textit{Notation:}
The sets of real numbers and real $n$-dimensional vectors are denoted by $\R$ and $\R^n$, respectively. Vector inequalities are interpreted componentwise. The symbols $\|\cdot\|$ and $\|\cdot\|_\infty$ denote the Euclidean
and infinity norms, respectively. For a vector $\mathbf v$,
$[\mathbf v]_+$ denotes its componentwise positive part. For a nonempty
set $\mathcal S$,
$\operatorname{dist}(\mathbf x,\mathcal S):=
\inf_{\mathbf y\in\mathcal S}\|\mathbf x-\mathbf y\|$. The symbols $\mathbf I$ and $\mathbf0$ denote the identity matrix and the zero vector or matrix of appropriate dimensions.

\section{Problem Formulation and Preliminaries}
\label{sec:problem}

In this section, we formulate the constrained optimization problem
and introduce the output-regulation and CBF methods.

\subsection{Constrained Optimization Problem}

Consider the constrained optimization problem
\begin{equation}
\label{eq:problem}
\begin{aligned}
\min_{\mathbf x\in\R^n}\quad & f(\mathbf x),\\
\operatorname{s.t.}\quad
& \mathbf h(\mathbf x)=\mathbf0,\\
& \mathbf g(\mathbf x)\leq\mathbf0,
\end{aligned}
\end{equation}
where $f:\R^n\to\R$, $\mathbf h:\R^n\to\R^p$, and
$\mathbf g:\R^n\to\R^m$ are twice continuously differentiable.
Define the inequality-feasible set and the feasible set, respectively, as
\[
\Sg:=\{\mathbf x\in\R^n:\mathbf g(\mathbf x)\leq\mathbf0\},
\
\mathcal F:=\{\mathbf x\in\Sg:\mathbf h(\mathbf x)=\mathbf0\}.
\]
The constraint Jacobians are
\(
\mathbf H(\mathbf x):=
[
\nabla h_1(\mathbf x)^\top,\dots,
\nabla h_p(\mathbf x)^\top
]^\top\in\R^{p\times n}\) and \(
\mathbf G(\mathbf x):=
[
\nabla g_1(\mathbf x)^\top,\dots,
\nabla g_m(\mathbf x)^\top
]^\top\in\R^{m\times n}.
\)

We introduce the
following assumptions for \eqref{eq:problem}.

\begin{assumption}[Equality constant rank]
\label{ass:B2}
The equality Jacobian $\mathbf H(\mathbf x)$ has constant rank on
$\mathcal U$. 
\end{assumption}

{Define the active inequality set and its Jacobian by
$\mathcal A^\star:=\{j:g_j(\mathbf x^\star)=0\}$ and
$\mathbf A^\star:=\mathbf G_{\mathcal A^\star}(\mathbf x^\star)$.
}

\begin{assumption}[Active-inequality LICQ]
\label{ass:B3}
{
The active inequality gradients are linearly independent, namely,
$\rank\mathbf A^\star=|\mathcal A^\star|$.
}
\end{assumption}

\begin{assumption}[Strict complementarity]
\label{ass:B4}
{
The inequality multiplier associated with the selected KKT triple satisfies
$\mu_j^\star>0$ for every $j\in\mathcal A^\star$.
}
\end{assumption}

{
Let $\mathbf H^\star:=\mathbf H(\mathbf x^\star)$ and
$\mathbf Q^\star:=\nabla_{\mathbf x\mathbf x}^2
\mathcal L(\mathbf x^\star,\mathbf z^\star,\boldsymbol{\mu}^\star)$.
}

\begin{assumption}[Curvature condition]
\label{ass:B5}
There exists a constant $\kappa\geq0$ such that
\(
\mathbf d^\top
\bigl(
\mathbf Q^\star
+\kappa\mathbf H^{\star\top}\mathbf H^\star
\bigr)
\mathbf d>0
\)
for every nonzero $\mathbf d\in\R^n$ satisfying
$\mathbf A^\star\mathbf d=\mathbf0$.
\end{assumption}

\begin{remark}
\label{rem:comparison_local_assumptions}
Assumptions~\ref{ass:B2}--\ref{ass:B5} are standard local conditions
in nonlinear programming and closely parallel those adopted in
\cite{allibhoy2024control}. The comparison with \cite{allibhoy2024control} concerns the
constraint qualifications used for particular convergence results.
That work analyzes a filter incorporating both equality and
inequality conditions and uses joint constraint qualifications.
In particular, some of its local and global convergence results
assume LICQ for the combined equality and active-inequality
gradients:
\begin{equation}
\label{eq:combined_licq}
\rank
\begin{bmatrix}
\mathbf H^\star\\
\mathbf A^\star
\end{bmatrix}
=p+|\mathcal A^\star|,
\end{equation}
which rules out linear dependence between the equality gradients and the
active inequality gradients. Assumptions~\ref{ass:B2} and \ref{ass:B3} instead impose regularity
on the two classes of constraints separately. They permit overlap
between the row spaces of $\mathbf H^\star$ and $\mathbf A^\star$.
The following example illustrates a configuration covered by our
local result in which the combined LICQ condition does not hold.
{This relaxation, which follows from the structure of the proposed ICBF-OR
flow, accommodates a broader class of problems. For example, consider
$f(\mathbf x)=-x_1$, $h(\mathbf x)=x_1$, and
$g(\mathbf x)=x_1+x_2^2$.
}
At $\mathbf x^\star=\mathbf0$, the multipliers
$z^\star=0$ and $\mu^\star=1$ form a KKT triple, and
\(
\mathbf H^\star
=
\mathbf A^\star
=
\begin{bmatrix}1&0\end{bmatrix}.
\)
Thus, Assumptions~\ref{ass:B2} and \ref{ass:B3} hold, whereas
{
\[
\rank
\begin{bmatrix}
\mathbf H^\star\\
\mathbf A^\star
\end{bmatrix}
=1<2=p+|\mathcal A^\star|.
\]
}
Hence, the combined LICQ condition \eqref{eq:combined_licq} fails.
\end{remark}

The Lagrangian associated with \eqref{eq:problem} is
\[
\mathcal L(\mathbf x,\mathbf z,\boldsymbol{\mu})
:=f(\mathbf x)+\mathbf z^\top\mathbf h(\mathbf x)
+\boldsymbol{\mu}^\top\mathbf g(\mathbf x),
\]
where $\mathbf z\in\R^p$ and $\boldsymbol{\mu}\in\R^m$ denote
the equality and inequality multipliers, respectively.
A point $\mathbf x^\star\in\mathcal F$ is a
Karush--Kuhn--Tucker (KKT) point of \eqref{eq:problem} if there
exist multipliers $\mathbf z^\star\in\R^p$ and
$\boldsymbol{\mu}^\star\in\R^m$ satisfying
\begin{subequations}
\label{eq:KKT}
\begin{align}
\nabla f(\mathbf x^\star)
+\mathbf H(\mathbf x^\star)^\top\mathbf z^\star
+\mathbf G(\mathbf x^\star)^\top\boldsymbol{\mu}^\star
&=\mathbf0,
\label{eq:KKTa}\\
\boldsymbol{\mu}^\star&\geq\mathbf0,
\label{eq:KKTb}\\
\mu_j^\star g_j(\mathbf x^\star)&=0,
\ j=1,\ldots,m.
\label{eq:KKTc}
\end{align}
\end{subequations}
Let $\KKT\subseteq\mathcal F$ denote the set of KKT points
of \eqref{eq:problem}.

\medskip
\noindent\textbf{Objective:}
Our objective is to design  optimization flows
whose primal state converges to a KKT point of \eqref{eq:problem}.

\subsection{Problem Reformulation}

We reformulate \eqref{eq:problem} from a control-theoretic perspective.
Consider the virtual control system
\begin{equation}
\label{eq:reformulated_system}
\dot{\mathbf x}
=-\nabla f(\mathbf x)+\mathbf u,
\qquad
\mathbf y=\mathbf h(\mathbf x),
\qquad
\mathbf x\in\Sg,
\end{equation}
where $\mathbf u\in\R^n$ is the control input. In this formulation,
$-\nabla f(\mathbf x)$ provides the nominal objective-descent direction,
the equality-constraint residual $\mathbf h(\mathbf x)$ is treated as the
regulated output, and the inequality constraints define the admissible
state set $\Sg$. Hence, \eqref{eq:problem} is recast as a
state-constrained output-regulation problem.

For \eqref{eq:reformulated_system}, the task is to design the control input
$\mathbf u$ such that $\mathbf x(t)\in\Sg$ for all $t\geq0$ whenever
$\mathbf x(0)\in\Sg$, and $\mathbf y(t)\to\mathbf0$ as $t\to\infty$,
thereby satisfying the inequality and equality constraints of
\eqref{eq:problem}, respectively. Meanwhile, $\mathbf u$ should ensure
that the closed-loop system converges to an equilibrium satisfying the KKT conditions
in \eqref{eq:KKT}, so that $\mathbf x(t)$ converges to a KKT point.
These requirements constitute the control-theoretic formulation of the
objective associated with \eqref{eq:problem}.

\begin{remark}
\label{rem:constraint_roles}
In the proposed framework, we treat equality and inequality constraints
separately. The practical rationale is that the two classes of constraints
often have distinct requirements along the optimization trajectory.
Equality constraints often represent steady-state requirements and may be
temporarily violated during transients, as occurs with auxiliary-variable
consistency and consensus constraints
\cite{Dhingra2019Proximal,Zeng2023Dynamical}. By contrast, inequality
constraints often encode hard physical or safety limits, such as
minimum-following-distance and state--input constraints
\cite{Ames2017Control,Wabersich2023Predictive,Rosolia2019Learning,
Taylor2020Adaptive}, and must therefore remain satisfied throughout the
trajectory.
\end{remark}

\subsection{Control Barrier Functions for Inequality Constraints}

Control barrier functions (CBFs) provide a systematic control-theoretic
framework for rendering a prescribed safe set forward invariant through
pointwise constraints on the system vector field
\cite{Ames2017Control}. {Consider the control system
$\dot{\mathbf x}=\mathbf v,$ where $\mathbf v\in\R^n$ is the control input.
Associate each inequality
in \eqref{eq:problem} with the barrier function
$b_j(\mathbf x):=-g_j(\mathbf x)$. A standard CBF condition is
$\nabla b_j(\mathbf x)^\top\mathbf v\geq
-\alpha_j(b_j(\mathbf x))$, where $\alpha_j$ is a locally Lipschitz
extended class-$\mathcal K$ function. A common
choice is $\alpha_j(s)=\gamma_js$ with $\gamma_j>0$.
}

To characterize the resulting trajectories, let $\eta_j(t;s)$ denote
the solution of
$\dot{\eta}_j=-\alpha_j(\eta_j)$ with $\eta_j(0;s)=s$. By the
comparison lemma, every trajectory satisfying the CBF condition obeys
{
\[
-g_j(\mathbf x(t))
\geq
\eta_j\bigl(t;-g_j(\mathbf x(0))\bigr).
\]
}
If $\mathbf x(0)\in\Sg$, then $-g_j(\mathbf x(0))\geq0$. Since
$\alpha_j(0)=0$ and $\alpha_j$ is strictly increasing,
$\eta_j(t;-g_j(\mathbf x(0)))$ remains nonnegative. Consequently,
$g_j(\mathbf x(t))\leq0$ for every $j$ throughout the interval of existence of the trajectory.

At any point satisfying $g_j(\mathbf x)=0$, one has
$\alpha_j(-g_j(\mathbf x))=0$, and the CBF condition reduces to
$\nabla g_j(\mathbf x)^\top\mathbf v\leq0$. Thus, independently of the
particular choice of $\alpha_j$, the CBF condition recovers the
standard non-outward-pointing condition on the boundary.

\section{{Inequality-CBF-Filtered Output--Regulation Flow}}
\label{sec:algorithm}

In this section, we develop the proposed Inequality-CBF-filtered Output--Regulation (ICBF-OR) flow.

Following the state-constrained output-regulation formulation in
Section~\ref{sec:problem}, we first construct a nominal output regulator
for the equality constraints. For the regulated output
$\mathbf y=\mathbf h(\mathbf x)$, consider the proportional--integral
feedback
\begin{equation}
\label{eq:equality_output_regulator}
\mathbf u_{\mathrm{eq}}
=-\mathbf H(\mathbf x)^\top
\bigl(\mathbf z+\mathbf K_p\mathbf y\bigr),
\qquad
\dot{\mathbf z}=\mathbf K_i\mathbf y,
\end{equation}
where $\mathbf K_p,\mathbf K_i\in\R^{p\times p}$ are symmetric positive
definite and $\mathbf z\in\R^p$ is the integral state. The proportional
term provides direct correction based on the equality residual, while
$\mathbf z$ accumulates this residual to provide integral action.
Substituting \eqref{eq:equality_output_regulator} into the nominal state
equation in \eqref{eq:reformulated_system} gives the output-regulated
direction
\begin{equation}
\label{eq:vpd}
\vpd(\mathbf x,\mathbf z)
:=-\nabla f(\mathbf x)-\mathbf H(\mathbf x)^\top\mathbf z
-\mathbf H(\mathbf x)^\top\mathbf K_p\mathbf h(\mathbf x).
\end{equation}
The inequality constraints are then enforced by filtering this nominal
direction. Collect the extended class-$\mathcal K$
functions $\alpha_j:\R\to\R$, $j=1,\ldots,m$, into the
componentwise mapping
$\boldsymbol{\alpha}(\mathbf s):=
(\alpha_1(s_1),\ldots,\alpha_m(s_m))^\top$
for $\mathbf s\in\R^m$.
For every $(\mathbf x,\mathbf z)\in\Sg\times\R^p$, define the filtered direction as the solution of the following quadratic program:
\begin{equation}
\label{eq:cbf_qp}
\begin{aligned}
\mathbf v^\star(\mathbf x,\mathbf z)
:=\underset{\mathbf v\in\R^n}{\arg\min}\quad
&\frac{1}{2}\left\|\mathbf v-\vpd(\mathbf x,\mathbf z)\right\|^2,\\
\operatorname{s.t.}\quad
&\mathbf G(\mathbf x)\mathbf v
-\boldsymbol{\alpha}\bigl(-\mathbf g(\mathbf x)\bigr)\leq\mathbf0.
\end{aligned}
\end{equation}
Thus, $\mathbf v^\star$ is the smallest Euclidean modification of the nominal direction $\vpd$ that satisfies all inequality CBF conditions. The proposed
ICBF-OR flow is
\begin{equation}
\label{eq:proposed_dynamics}
\begin{aligned}
\dot{\mathbf x}&=\mathbf v^\star(\mathbf x,\mathbf z),\\
\dot{\mathbf z}&=\mathbf K_i\mathbf h(\mathbf x),
\end{aligned}
\end{equation}
where $\mathbf v^\star(\mathbf x,\mathbf z)$ is computed at each state by solving the embedded CBF quadratic program \eqref{eq:cbf_qp}.

Next, we discuss the solving methods of 
the CBF quadratic program \eqref{eq:cbf_qp}. It is linearly constrained and
strongly convex, and therefore admits a unique primal optimizer for every
$(\mathbf x,\mathbf z)\in\Sg\times\R^p$. In favorable cases, this optimizer
can be evaluated algebraically using the classical result in
\cite{Tondel2003Algorithm}.
Suppose the rows of $\mathbf G_{\mathcal I}(\mathbf x)$ are linearly
independent for $\mathcal I$ in \eqref{eq:global_qp_active_set}. Then the optimizer
and its multiplier are given by
\[
\begin{aligned}
\boldsymbol{\lambda}_{\mathcal I}^\star
&:=\left(\mathbf G_{\mathcal I}(\mathbf x)
\mathbf G_{\mathcal I}(\mathbf x)^{\top}\right)^{-1}
\left[
\mathbf G(\mathbf x)\vpd(\mathbf x,\mathbf z)
-\boldsymbol{\alpha}(-\mathbf g(\mathbf x))
\right]_{\mathcal I},
\\
\boldsymbol{\lambda}_{\mathcal I^c}^\star&:=\mathbf0,\\
\mathbf v^\star
&:=\vpd(\mathbf x,\mathbf z)-\mathbf G_{\mathcal I}(\mathbf x)^{\top}
\boldsymbol{\lambda}_{\mathcal I}^\star.
\end{aligned}
\]

When the binding set is unknown or its gradients are linearly dependent,
an exact optimizer can instead be obtained through an inner iterative
solver. The output--regulation method in \cite{Patrinos2014Accelerated} gives the
projected dual-gradient flow
\begin{subequations}
\label{eq:exact_qp_iteration}
\begin{align}
\tau\dot{\boldsymbol{\lambda}}
&=
\left[
\mathbf G(\mathbf x)\vpd(\mathbf x,\mathbf z)
-\boldsymbol{\alpha}\bigl(-\mathbf g(\mathbf x)\bigr)
-\mathbf G(\mathbf x)\mathbf G(\mathbf x)^\top\boldsymbol{\lambda}
\right]_{\boldsymbol{\lambda}}^{+},
\label{eq:exact_qp_continuous_multiplier}\\
\mathbf v
&=
\vpd(\mathbf x,\mathbf z)-\mathbf G(\mathbf x)^\top\boldsymbol{\lambda},
\label{eq:exact_qp_continuous_primal}
\end{align}
\end{subequations}
where, for each $j\in\{1,\ldots,m\}$,
\[
[w_j]^+_{\lambda_j}
:=
\begin{cases}
w_j, & \lambda_j>0,\\
\max\{w_j,0\}, & \lambda_j=0,
\end{cases}
\]
$\boldsymbol{\lambda}(0)\geq\mathbf0$, and $\tau>0$ is the inner time
constant. The outer variables $(\mathbf x,\mathbf z)$ in \eqref{eq:proposed_dynamics} are held fixed when
\eqref{eq:exact_qp_iteration} is integrated. By
\cite{Patrinos2014Accelerated},
$\mathbf v\to\mathbf v^\star(\mathbf x,\mathbf z)$ for \eqref{eq:exact_qp_iteration}, which, however, incurs a substantially higher computational cost per evaluation than using the analytical solution.

% \begin{remark}
% \label{rem:design_motivation}
% The ICBF-OR flow is inspired by the augmented output--regulation dynamics in
% \cite{Tang2020SemiGlobal} and the CBF-based safe gradient flow in
% \cite{allibhoy2024control}. The former handles equality and inequality
% constraints within a unified output--regulation framework but does not generally
% preserve inequality feasibility during transients. Such feasibility is
% essential when the inequalities encode hard physical or safety limits
% \cite{Ames2017Control,Wabersich2023Predictive}. The latter enforces both
% types of constraints through CBF-based velocity conditions and thereby
% renders the feasible set forward invariant. However, as noted in
% \cite{allibhoy2024control}, feasibility of the resulting mixed
% equality--inequality CBF program can be difficult to guarantee outside the
% feasible set. Motivated by these complementary features, the proposed flow
% employs a proportional--integral output regulator for the equality constraints and a CBF
% filter only for the inequalities. This separation makes the CBF quadratic program feasible for every
% $\mathbf x\in\Sg$ and renders $\Sg$ forward invariant, as established in
% Propositions~\ref{prop:qp_feasibility} and
% \ref{prop:forward_invariance}, respectively.
% \end{remark}

\section{Feasibility, Safety, and Convergence Analysis}
\label{sec:convergence}

In this section, we first establish feasibility of the CBF quadratic program and forward invariance of $\Sg$ under the ICBF-OR flow, and then analyze its local exponential convergence and convergence from an entire connected component of the inequality-feasible set.

\subsection{Feasibility and Forward Invariance}

We establish that the CBF quadratic program remains feasible throughout $\Sg$ and that $\Sg$ is forward invariant under \eqref{eq:proposed_dynamics}.

\begin{proposition}[Feasibility of the CBF quadratic program]
\label{prop:qp_feasibility}
For every \(\mathbf x\in\Sg\) and every \(\mathbf z\in\R^p\), the quadratic program
\eqref{eq:cbf_qp} is feasible and admits a unique optimizer. 
\end{proposition}

\begin{proposition}[Forward invariance of the inequality-feasible set]
\label{prop:forward_invariance}
Let \((\mathbf x(t),\mathbf z(t))\) be any solution of
\eqref{eq:proposed_dynamics} with \(\mathbf x(0)\in\Sg\). Then, 
\begin{equation}
\label{eq:constraint_comparison}
g_j(\mathbf x(t))
\leq-\eta_j\bigl(t;-g_j(\mathbf x(0))\bigr)
\leq0,
\qquad j=1,\ldots,m.
\end{equation}
\end{proposition}

\subsection{Local Convergence}
\label{subsec:local_convergence}

We next establish the local convergence of the ICBF-OR flow around a
given KKT point. 
Fix a KKT triple
$(\mathbf x^\star,\mathbf z^\star,\boldsymbol{\mu}^\star)$ and let
$\mathcal U$ be a neighborhood of $\mathbf x^\star$. For the remaining
analysis, we additionally assume that each extended class-$\mathcal K$
function $\alpha_j$ is continuously differentiable and satisfies
$\alpha_j'(s)>0$ for all $s\geq0$ and $j=1,\ldots,m$.
In particular, the standard linear choice
$\alpha_j(s)=\gamma_j s$ satisfies this condition.

First, we introduce the following lemma to show equilibrium–KKT equivalence of \eqref{eq:proposed_dynamics}.

\begin{lemma}[Equilibrium--KKT equivalence]
\label{lem:equilibrium_KKT}
A pair $(\mathbf x^\star,\mathbf z^\star)\in\Sg\times\R^p$ is an equilibrium of
\eqref{eq:proposed_dynamics} if and only if there exists
$\boldsymbol{\mu}^\star\in\R_{\geq0}^m$ such that
$(\mathbf x^\star,\mathbf z^\star,\boldsymbol{\mu}^\star)$ is a KKT triple of
\eqref{eq:problem}. 
\end{lemma}

Since we do not impose the combined LICQ
condition, the equality and active-inequality gradients may be
linearly dependent, and the same primal KKT point $\mathbf x^\star$ may
admit multiple equality and inequality multiplier pairs. The following
lemma characterizes all such multiplier pairs.

\begin{lemma}[KKT multipliers at $\mathbf x^\star$]
\label{lem:local_kkt_set}
Suppose Assumption~\ref{ass:B3} holds. Let
$\mathbf N^\star$ be a matrix whose columns form an orthonormal basis of
$\ker\mathbf A^\star$, and define
\(
\mathbf B^\star:=\mathbf H^\star\mathbf N^\star.
\)
For every $\boldsymbol{\eta}\in\R^p$, define
\begin{equation}
\label{eq:local_inequality_multiplier}
\begin{aligned}
\boldsymbol{\mu}_{\mathcal A^\star}(\boldsymbol{\eta})
&:=
\boldsymbol{\mu}_{\mathcal A^\star}^\star
-
\left(
\mathbf A^\star\mathbf A^{\star\top}
\right)^{-1}
\mathbf A^\star\mathbf H^{\star\top}\boldsymbol{\eta},\\
\boldsymbol{\mu}_{(\mathcal A^\star)^c}(\boldsymbol{\eta})
&:=\mathbf0,
\end{aligned}
\end{equation}
and
\[
\mathcal M^\star
:=
\left\{
\boldsymbol{\eta}\in\R^p:
\mathbf B^{\star\top}\boldsymbol{\eta}=\mathbf0,\ 
\boldsymbol{\mu}_{\mathcal A^\star}(\boldsymbol{\eta})\geq\mathbf0
\right\}.
\]
Then $(\mathbf x^\star,\mathbf z,\boldsymbol{\mu})$ is a KKT triple of
\eqref{eq:problem} if and only if
$\mathbf z=\mathbf z^\star+\boldsymbol{\eta}$ and
$\boldsymbol{\mu}=\boldsymbol{\mu}(\boldsymbol{\eta})$ for some
$\boldsymbol{\eta}\in\mathcal M^\star$.
\end{lemma}

Since the state of the ICBF-OR flow consists of
$(\mathbf x,\mathbf z)$, define
\begin{equation}
\label{eq:local_equilibrium_set}
\mathcal E^\star
:=
\left\{
\left(\mathbf x^\star,\mathbf z^\star+\boldsymbol{\eta}\right):
\boldsymbol{\eta}\in\mathcal M^\star
\right\}.
\end{equation}

By Lemmas~\ref{lem:equilibrium_KKT} and~\ref{lem:local_kkt_set}, $\mathcal E^\star$ is exactly the
set of equilibria of the ICBF-OR flow whose primal component is
$\mathbf x^\star$. The following theorem shows that, for initial states in $\Sg\times\R^p$ sufficiently close to $(\mathbf x^\star,\mathbf z^\star)$, the primal state $\mathbf x(t)$ of the ICBF-OR flow converges exponentially to the KKT point $\mathbf x^\star$.
\begin{theorem}[Local exponential convergence]
\label{thm:local_convergence}
Suppose Assumptions~\ref{ass:B2}--\ref{ass:B5} hold. Choose
$\mathbf K_p=k_p\mathbf I$ with $k_p>\kappa$. Let
$\boldsymbol\xi:=(\mathbf x,\mathbf z)$ and
$\boldsymbol\xi^\star:=(\mathbf x^\star,\mathbf z^\star)$. Then there exist
a neighborhood $\mathcal W$ of $\boldsymbol\xi^\star$ and constants
$c,\beta>0$ such that every solution of \eqref{eq:proposed_dynamics} with
$\boldsymbol\xi(0)\in\mathcal W\cap(\Sg\times\R^p)$ converges to some
$\bar{\boldsymbol\xi}=(\mathbf x^\star,\bar{\mathbf z})\in\mathcal E^\star$
and satisfies
\begin{equation}
\label{eq:local_exp_estimate}
\|\boldsymbol\xi(t)-\bar{\boldsymbol\xi}\|
\leq c\,e^{-\beta t}
\operatorname{dist}\bigl(\boldsymbol\xi(0),\mathcal E^\star\bigr),
\qquad t\geq0.
\end{equation}
\end{theorem}

\subsection{Convergence in the Inequality-Feasible Set}
\label{subsec:safe_set_convergence}

Theorem~\ref{thm:local_convergence} establishes local exponential
convergence around a prescribed KKT point. Since the proposed ICBF-OR flow is well-defined throughout $\Sg$ owing to the persistent feasibility of its CBF quadratic program, as established in Proposition~\ref{prop:qp_feasibility}, its attraction region can potentially be enlarged to $\Sg$. We next provide conditions under
which the primal attraction region contains an entire connected component of
the inequality-feasible set $\Sg$. 

Fix a KKT triple
$(\mathbf x^\star,\mathbf z^\star,\boldsymbol{\mu}^\star)$. Let
$\Sg^\star$ be the connected component of $\Sg$ containing
$\mathbf x^\star$. {We assume that the feasible subset
$\mathcal F^\star:=\mathcal F\cap\Sg^\star$ is compact.}
 For
each $(\mathbf x,\mathbf z)\in\Sg^\star\times\R^p$, let
$\boldsymbol{\lambda}^\star(\mathbf x,\mathbf z)$ denote the multiplier of the
CBF constraint in \eqref{eq:cbf_qp}, and define its binding set by
\begin{equation}
\label{eq:global_qp_active_set}
\mathcal I(\mathbf x,\mathbf z)
:=
\left\{
j:
\nabla g_j(\mathbf x)^\top\mathbf v^\star(\mathbf x,\mathbf z)
-\alpha_j\bigl(-g_j(\mathbf x)\bigr)=0
\right\}.
\end{equation}

\begin{assumption}[Equality uniform regularity and active-inequality LICQ]
\label{ass:C1}
{
After
redundant equality equations are removed, there exists $\sigma_h>0$ such
that $\mathbf H(\mathbf x)\mathbf H(\mathbf x)^\top
\succeq\sigma_h^2\mathbf I$ for every $\mathbf x\in\Sg^\star$.
Moreover, for every binding set in \eqref{eq:global_qp_active_set}, the rows
of $\mathbf G_{\mathcal I}(\mathbf x)$ are linearly independent, uniformly
on $\Sg^\star\times\R^p$.
}
\end{assumption}

% \begin{assumption}[Strict complementarity]
% \label{ass:C2}
% {
% Every CBF constraint that is binding in the interior of $\Sg^\star$
% has a positive QP multiplier, i.e., $j\in\mathcal I(\mathbf x,\mathbf z)$
% and $g_j(\mathbf x)<0$ imply
% $\lambda_j^\star(\mathbf x,\mathbf z)>0$.
% }
% \end{assumption}

For later use, define
\begin{equation}
\label{eq:global_lagrangian_hessian}
\mathbf Q(\mathbf x,\mathbf z)
:=
\nabla_{\mathbf x\mathbf x}^2
\mathcal L\!\left(
\mathbf x,
\mathbf z+\mathbf K_p\mathbf h(\mathbf x),
\boldsymbol{\lambda}^\star(\mathbf x,\mathbf z)
\right).
\end{equation}

\begin{assumption}[Curvature condition]
\label{ass:A8}
There exists a constant $\kappa\geq0$ such that, for every
$(\mathbf x,\mathbf z)\in\Sg^\star\times\R^p$,
\begin{equation}
\label{eq:global_directional_curvature}
\mathbf v^{\star\top}
\bigl(
\mathbf Q(\mathbf x,\mathbf z)
+\kappa\mathbf H(\mathbf x)^\top\mathbf H(\mathbf x)
\bigr)
\mathbf v^\star
\geq0.
\end{equation}
\end{assumption}

{For every $\mathbf x\in\Sg^\star$, let
$\mathcal A(\mathbf x):=\{j:g_j(\mathbf x)=0\}$.
}

\begin{assumption}[Boundary-compatible equality restoration]
\label{ass:A6}
{
The equality-restoration direction is compatible with the active
inequality constraints, namely,
$
\mathbf G_{\mathcal A(\mathbf x)}(\mathbf x)
\mathbf H(\mathbf x)^\top\mathbf h(\mathbf x)
\geq\mathbf0$.
}
\end{assumption}

\begin{assumption}[Mixed MFCQ]
\label{ass:global_mfcq}
{
For every $\mathbf x\in\Sg^\star$, there exists a direction
$\mathbf d_{\mathbf x}\in\R^n$ such that
$\mathbf H(\mathbf x)\mathbf d_{\mathbf x}=\mathbf0$ and
$\mathbf G_{\mathcal A(\mathbf x)}(\mathbf x)
\mathbf d_{\mathbf x}<\mathbf0$.
}
\end{assumption}

\begin{remark}
\label{rem:global_assumptions}
Assumption~\ref{ass:C1} is the global counterpart of
Assumptions~\ref{ass:B2} and~\ref{ass:B3}.
Assumption~\ref{ass:A8} can be viewed as a global extension of
Assumption~\ref{ass:B5}, replacing the active-inequality tangent
directions $\mathbf d$ satisfying $\mathbf A^\star\mathbf d=\mathbf0$
at the selected KKT point with the actual direction
$\mathbf v^\star$ throughout $\Sg^\star\times\R^p$. It is
automatically satisfied with $\kappa=0$ for convex programs.
Assumption~\ref{ass:A6} ensures that the equality-restoration direction
$-\mathbf H(\mathbf x)^\top\mathbf h(\mathbf x)$ does not point outside
$\Sg^\star$ at its boundary. Assumption~\ref{ass:global_mfcq}
is the standard MFCQ and is commonly adopted in convergence
analyses of constrained optimization
\cite{Boob2023Stochastic,Jia2025FirstOrder}.
\end{remark}

The following result shows that, under the preceding assumptions, the
 attraction region of the ICBF-OR flow contains the entire connected
component $\Sg^\star$.

\begin{theorem}[Safe-component convergence]
\label{thm:safe_component_convergence}
Suppose Assumptions~\ref{ass:C1}--\ref{ass:global_mfcq} hold and $\mathbf z(0)=\mathbf0$.
Choose $\mathbf K_p=k_p\mathbf I$ with $k_p>\kappa$ and
$\mathbf K_i=k_i\mathbf I$ with $k_i>0$.
Then every solution of
\eqref{eq:proposed_dynamics} with $\mathbf x(0)\in\Sg^\star$ satisfies
\begin{equation}
\label{eq:global_convergence_conclusion}
\operatorname{dist}
\bigl(\mathbf x(t),\KKT\cap\Sg^\star\bigr)
\longrightarrow0.
\end{equation}
\end{theorem}

\section{The Approximate Implementation of the ICBF-OR Flow}
\label{sec:cbf_qp_solution}

The ICBF-OR flow requires solving the embedded CBF quadratic program
\eqref{eq:cbf_qp} at each state to obtain $\mathbf v^\star$. If a
closed-form solution is unavailable and the CBF quadratic program must instead be solved
iteratively, the inner solve can impose a substantial computational burden
on each evaluation of the flow. To address this issue, an alternative is to evolve the quadratic-program
multiplier concurrently with $(\mathbf x,\mathbf z)$. This gives an
approximate implementation of the ICBF-OR \eqref{eq:proposed_dynamics}.

{
Instead of solving \eqref{eq:cbf_qp} exactly for each
$(\mathbf x,\mathbf z)$, we embed
\eqref{eq:exact_qp_continuous_multiplier} into
\eqref{eq:proposed_dynamics}, replacing
$\mathbf v^\star(\mathbf x,\mathbf z)$ with the value of $\mathbf v$ in
\eqref{eq:exact_qp_continuous_primal}. The resulting approximate ICBF-OR
(A-ICBF-OR) flow is
\begin{subequations}
\label{eq:dynamic_icbf_pd}
\begin{align}
\dot{\mathbf x}
={}&
-\nabla f(\mathbf x)
-\mathbf H(\mathbf x)^\top\mathbf z
-\mathbf H(\mathbf x)^\top\mathbf K_p\mathbf h(\mathbf x)\nonumber
\\&-\mathbf G(\mathbf x)^\top\boldsymbol{\lambda},
\label{eq:dynamic_icbf_pd_primal}\\
\dot{\mathbf z}
={}&
\mathbf K_i\mathbf h(\mathbf x),
\label{eq:dynamic_icbf_pd_equality}\\
\tau\dot{\boldsymbol{\lambda}}
={}&
\Big[
\mathbf G(\mathbf x)\dot{\mathbf x}
-\boldsymbol{\alpha}\bigl(-\mathbf g(\mathbf x)\bigr)
\Big]^+_{\boldsymbol{\lambda}}.
\label{eq:dynamic_icbf_pd_inequality}
\end{align}
\end{subequations}
Here, $\boldsymbol{\lambda}(0)\geq\mathbf0$ and $\dot{\mathbf x}$ in
\eqref{eq:dynamic_icbf_pd_inequality} can be replaced by the right-hand
side of \eqref{eq:dynamic_icbf_pd_primal}; $\dot{\mathbf x}$ is retained
for compactness.
}

The following result is the counterpart of
Lemma~\ref{lem:equilibrium_KKT} for the A-ICBF-OR flow.

\begin{lemma}[Equilibrium--KKT equivalence]
\label{lem:dynamic_equilibrium_KKT}
A triple
\(
(\mathbf x^\star,\mathbf z^\star,\boldsymbol{\lambda}^\star)
\in
\R^n\times\R^p\times\R_{\geq0}^m
\)
is an equilibrium of the A-ICBF-OR flow
\eqref{eq:dynamic_icbf_pd} if and only if it is a KKT triple of
\eqref{eq:problem}, with $\boldsymbol{\mu}^\star=\boldsymbol{\lambda}^\star$ being the
inequality multiplier.
\end{lemma}

Analogously to \eqref{eq:local_equilibrium_set}, define
\begin{equation}
\label{eq:dynamic_local_equilibrium_set}
\mathcal E_{\mathrm D}^\star
:=
\left\{
\left(
\mathbf x^\star,
\mathbf z^\star+\boldsymbol{\eta},
\boldsymbol{\mu}(\boldsymbol{\eta})
\right):
\boldsymbol{\eta}\in\mathcal M^\star
\right\},
\end{equation}
where $\boldsymbol{\mu}(\boldsymbol{\eta})$ is defined in
\eqref{eq:local_inequality_multiplier}.
By Lemmas~\ref{lem:local_kkt_set} and~\ref{lem:dynamic_equilibrium_KKT},
$\mathcal E_{\mathrm D}^\star$ is exactly the set of equilibria of
\eqref{eq:dynamic_icbf_pd} whose primal component is $\mathbf x^\star$.
The following result is the dynamic counterpart of
Theorem~\ref{thm:local_convergence}.

\begin{theorem}[Local exponential convergence of the A-ICBF-OR flow]
\label{thm:dynamic_icbf_pd_convergence}
Suppose Assumptions~\ref{ass:B2}--\ref{ass:B5} hold. Choose
$\mathbf K_p=k_p\mathbf I$ with $k_p>\kappa$. Then there exists
$\tau^\ast>0$ such that, for every $\tau\in(0,\tau^\ast]$, there exist a
neighborhood $\mathcal W_\tau$ of
$\boldsymbol\xi_{\mathrm D}^\star
:=(\mathbf x^\star,\mathbf z^\star,\boldsymbol{\mu}^\star)$
and constants $c_\tau,\beta_\tau>0$ such that every solution of
\eqref{eq:dynamic_icbf_pd} with
$\boldsymbol\xi(0):=(\mathbf x(0),\mathbf z(0),\boldsymbol\lambda(0))
\in\mathcal W_\tau\cap(\R^n\times\R^p\times\R^m_{\geq0})$
converges to some
$\bar{\boldsymbol\xi}=(\mathbf x^\star,\bar{\mathbf z},\bar{\boldsymbol\lambda})\in\mathcal E_{\mathrm D}^\star$ and satisfies
\begin{equation}
\label{eq:dynamic_local_exp_estimate}
\|\boldsymbol\xi(t)-\bar{\boldsymbol\xi}\|
\leq c_\tau e^{-\beta_\tau t}
\operatorname{dist}\bigl(\boldsymbol\xi(0),\mathcal E_{\mathrm D}^\star\bigr),
\qquad t\geq0.
\end{equation}
\end{theorem}

The following result establishes an $O(\tau)$ bound on the transient
inequality-constraint violation of the A-ICBF-OR flow.

\begin{proposition}
[Transient inequality-constraint violation]
\label{prop:dynamic_constraint_violation}
Under the conditions of Theorem~\ref{thm:dynamic_icbf_pd_convergence},
there exists a constant $M>0$, independent of $\tau$, such that,
for every $\tau\in(0,\tau^\ast]$, there exists a neighborhood
$\mathcal U_\tau$ of $\boldsymbol\xi_{\mathrm D}^\star$
for which every solution of \eqref{eq:dynamic_icbf_pd} starting from
$\mathcal U_\tau\cap(\R^n\times\R^p\times\R^m_{\geq0})$ with
$\mathbf x(0)\in\Sg$ satisfies
\begin{equation}
\label{eq:dynamic_uniform_violation}
\sup_{t\geq0}
\bigl\|[\mathbf g(\mathbf x(t))]_+\bigr\|_\infty
\leq M\tau.
\end{equation}
\end{proposition}

Proposition~\ref{prop:dynamic_constraint_violation} quantifies the effect
of $\tau$ on transient inequality feasibility, showing that
the maximum constraint violation is at most $M\tau$.
To preserve the original inequality constraints, one may prescribe
a constraint-tightening margin $\varepsilon>0$ and replace each
$g_j$ in \eqref{eq:dynamic_icbf_pd} with
$\widetilde g_j:=g_j+\varepsilon$, provided that the tightened
problem remains feasible. By 
Proposition~\ref{prop:dynamic_constraint_violation}, we have 
$g_j(\mathbf x(t))\leq\tau\lambda_j(t)-\varepsilon
\leq M\tau-\varepsilon$.
Thus, choosing $0<\tau\leq\min\{\tau^\ast,\varepsilon/M\}$
ensures $\mathbf g(\mathbf x(t))\leq\mathbf0$ for all $t\geq0$ and therefore guarantees the forward invariance property.

From the perspective of per-evaluation computational cost, A-ICBF-OR is especially suitable when the ICBF-OR quadratic program does not admit a tractable closed-form solution. In this case, its explicit multiplier implementation avoids an inner iterative quadratic-program solve at each vector-field evaluation and can therefore reduce the empirical per-evaluation cost, as demonstrated in Section~\ref{subsubsec:flow_comparison}.

\section{Discussion: Relations Among the Optimization Flows}
\label{subsec:relations_solution_methods}

We next clarify that \eqref{eq:dynamic_icbf_pd}  provides a general parameterized formulation. It defines the A-ICBF-OR flow for $0<\tau<\infty$, recovers the ICBF-OR flow \eqref{eq:proposed_dynamics} as $\tau\to0^+$, and reduces to the projected primal--dual (P-PD) flow in the literature as $\tau\to\infty$.

The case $\tau=0$ in \eqref{eq:dynamic_icbf_pd} is understood as the algebraic fast-solver limit,
rather than as an ordinary differential equation. In this limit,
\eqref{eq:dynamic_icbf_pd_inequality} reduces to
\begin{equation}
\label{eq:dynamic_zero_tau_condition}
\left[
\mathbf G(\mathbf x)\dot{\mathbf x}
-\boldsymbol{\alpha}\bigl(-\mathbf g(\mathbf x)\bigr)
\right]^+_{\boldsymbol{\lambda}}
=\mathbf0.
\end{equation}
By the definition of the projected operator,
\eqref{eq:dynamic_zero_tau_condition} is equivalent to
\begin{align*}
\mathbf G(\mathbf x)\dot{\mathbf x}
-\boldsymbol{\alpha}\bigl(-\mathbf g(\mathbf x)\bigr)
&\leq\mathbf0,\\
\boldsymbol{\lambda}&\geq\mathbf0,\\
\lambda_j
\left[
\nabla g_j(\mathbf x)^\top\dot{\mathbf x}
-\alpha_j\bigl(-g_j(\mathbf x)\bigr)
\right]
&=0,
\qquad j=1,\ldots,m.
\end{align*}
Together with \eqref{eq:dynamic_icbf_pd_primal}, these are precisely
the KKT conditions of the CBF quadratic program \eqref{eq:cbf_qp}.
Since \eqref{eq:cbf_qp} has a unique primal optimizer, it follows that
\(
\dot{\mathbf x}
=
\mathbf v^\star(\mathbf x,\mathbf z).
\)
Consequently, the algebraic limit $\tau=0$ recovers the exact ICBF-OR flow \eqref{eq:proposed_dynamics}.

Moreover, choose $\boldsymbol{\alpha}(\mathbf s)$ by
\(
\boldsymbol{\alpha}_\tau(\mathbf s)
:=
\tau\mathbf s.
\)
Then \eqref{eq:dynamic_icbf_pd_inequality} becomes
\(
\dot{\boldsymbol{\lambda}}
=
\left[
\mathbf g(\mathbf x)
+\frac{1}{\tau}
\mathbf G(\mathbf x)\dot{\mathbf x}
\right]^+_{\boldsymbol{\lambda}}.
\)
Letting $\tau\to\infty$ gives
\(
\dot{\boldsymbol{\lambda}}
=
[\mathbf g(\mathbf x)]^+_{\boldsymbol{\lambda}}.
\)
Together with
\eqref{eq:dynamic_icbf_pd_primal} and
\eqref{eq:dynamic_icbf_pd_equality}, the limiting dynamics are
\begin{equation}
\label{eq:projected_primal_dual_flow}
\begin{aligned}
\dot{\mathbf x}
={}&
-\nabla f(\mathbf x)
-\mathbf H(\mathbf x)^\top\mathbf z
-\mathbf H(\mathbf x)^\top\mathbf K_p\mathbf h(\mathbf x)
-\mathbf G(\mathbf x)^\top\boldsymbol{\lambda},
\\
\dot{\mathbf z}
={}&
\mathbf K_i\mathbf h(\mathbf x),
\\
\dot{\boldsymbol{\lambda}}
={}&
[\mathbf g(\mathbf x)]^+_{\boldsymbol{\lambda}},
\end{aligned}
\end{equation}
which is the P-PD flow associated with
\eqref{eq:problem}. {Its convergence analysis is given in
\cite{Tang2020SemiGlobal}.}

The relations among the flows are summarized in
Table~\ref{tab:flow_relations}.

\begin{table}[t]
\centering
\caption{Flows recovered from the general formulation
\eqref{eq:dynamic_icbf_pd}.}
\label{tab:flow_relations}
 \footnotesize
    \setlength{\tabcolsep}{5pt}
    \begin{tabular}{lc}
\hline
\textbf{Time-scale regime} & \textbf{Resulting flow} \\
\hline
$0<\tau<\infty$
&
A-ICBF-OR flow \eqref{eq:dynamic_icbf_pd}
\\
$\tau\to0^+$
&
ICBF-OR flow \eqref{eq:proposed_dynamics}
\\
$\tau\to\infty$
&
P-PD flow
\eqref{eq:projected_primal_dual_flow}
\\
\hline
\end{tabular}
\end{table}

\section{Numerical Simulations}
\label{sec:numerical_simulations}

In this section, we provide numerical simulations to illustrate the
convergence and safety properties of the proposed flows. The extended class-$\mathcal K$ functions are selected as
$\alpha_j(s)=s$.

\subsection{Local and Safe-Component Convergence}

\begin{figure}[t]
\centering
\includegraphics[width=\linewidth]
{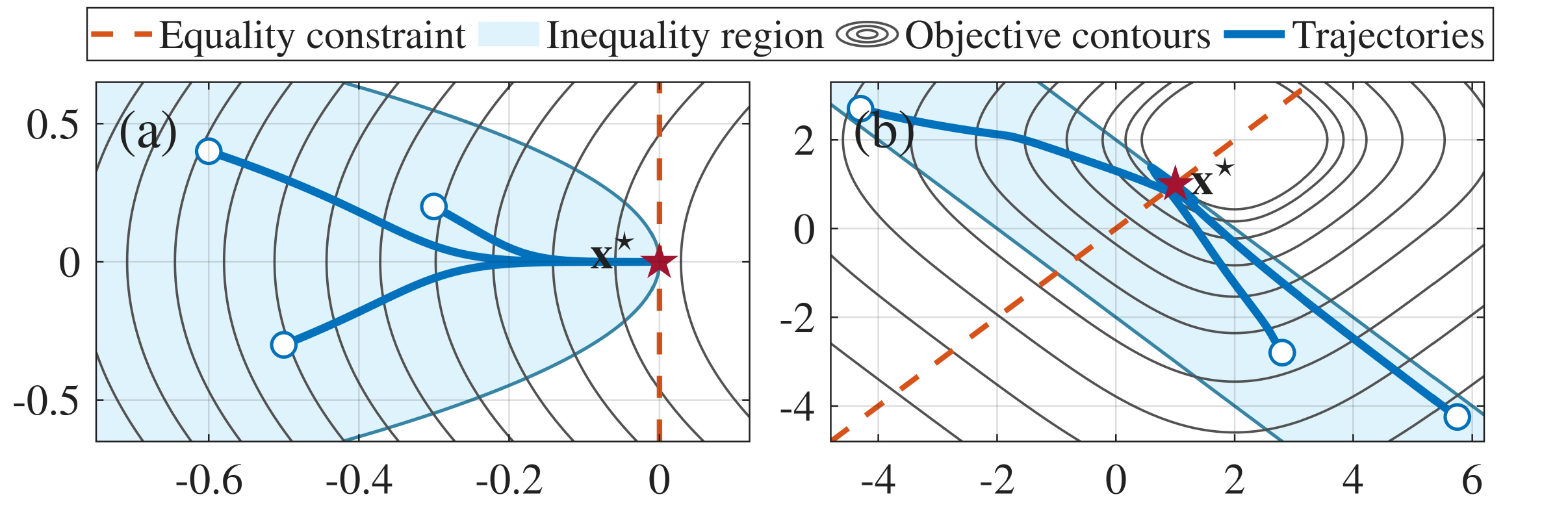}
\caption{Two-dimensional illustrations using the ICBF-OR flow. (a) Case I; (b) Case II.}
\label{fig:two_dimensional_examples}
\end{figure}

Two examples are provided below.
The first example illustrates our local convergence result when
the combined LICQ condition used for certain results in
\cite{allibhoy2024control} fails. The second illustrates the
safe-component convergence result.

\subsubsection{Case I: Local convergence}

Consider the nonlinear optimization problem
\[
\begin{aligned}
\min_{\mathbf x\in\mathbb R^2}\quad
& f(\mathbf x)
 =\frac{1}{2}(x_1-1.8)^2+\frac{1}{2}x_2^2,\\
\operatorname{s.t.}\quad
& {h(\mathbf x)=x_1=0,}\\
& {g(\mathbf x)=x_1+x_2^2\leq0.}
\end{aligned}
\]
The inequality-feasible set is
{
$\Sg=\{\mathbf x:x_1+x_2^2\leq0\}$.}
The selected KKT triple is
{
$\mathbf x^\star=\mathbf0$, $z^\star=0$, and
$\mu^\star=1.8$.}
At this KKT point,
{
$\mathbf H^\star=\begin{bmatrix}1&0\end{bmatrix}$ and
$\mathbf A^\star=\begin{bmatrix}1&0\end{bmatrix}$.}
{
Thus, the equality and active-inequality gradients are individually
linearly independent, whereas
\[
\rank
\begin{bmatrix}
\mathbf H^\star\\
\mathbf A^\star
\end{bmatrix}
=1<2,
\]
so the combined LICQ condition \eqref{eq:combined_licq} does not hold.}
The Lagrangian Hessian is
\(
\mathbf Q^\star
=
\nabla_{\mathbf x\mathbf x}^2
\mathcal L(\mathbf x^\star,z^\star,\mu^\star)
=
{
\begin{bmatrix}
1&0\\
0&4.6
\end{bmatrix}}
\succ0.
\)
Therefore, Assumptions~\ref{ass:B2}--\ref{ass:B5} hold locally around
$\mathbf x^\star$.

The ICBF-OR flow is implemented with $K_p=K_i=1$. Three initial
conditions in $\Sg$ are chosen close to the selected equilibrium. The
resulting state trajectories are shown in
Fig.~\ref{fig:two_dimensional_examples}(a). Although the initial
conditions do not satisfy the equality constraint, all trajectories remain in
$\Sg$ and converge to $\mathbf x^\star$. This behavior is consistent with the
local exponential convergence established in
Theorem~\ref{thm:local_convergence}.

\subsubsection{Case II: Convergence in a safe component}

Define the rotated coordinates
$s:=(x_1+x_2)/2$ and $r:=(x_1-x_2)/2$. We next consider the problem
\[
\begin{aligned}
\min_{\mathbf x\in\mathbb R^2}\quad
&f(\mathbf x)=
\frac{1}{4}r^4+\frac{1}{2}r^2
+\frac{1}{4}(s-2)^4+\frac{1}{2}(s-2)^2,\\
\operatorname{s.t.}\quad
&h(\mathbf x)=r=0,\qquad g(\mathbf x)=s^2-1\leq0.
\end{aligned}
\]
The inequality-feasible set is the closed and connected strip
\(
\Sg^\star
=
\left\{
\mathbf x:
|x_1+x_2|\leq2
\right\},
\)
which is noncompact, while
$\mathcal F^\star=\mathcal F\cap\Sg^\star
=\{\operatorname{col}(q,q):q\in[-1,1]\}$ is compact.
The unique KKT triple in this component is
$\mathbf x^\star=\operatorname{col}(1,1)$, $z^\star=0$, and
$\mu^\star=1$.

The assumptions of Theorem~\ref{thm:safe_component_convergence} can also
be verified explicitly. In particular,
\(
\mathbf H(\mathbf x)
=
\begin{bmatrix}
1/2&-1/2
\end{bmatrix},\
\mathbf G(\mathbf x)
=
\begin{bmatrix}
s&s
\end{bmatrix}.
\)
Because the equality-restoration and equality-multiplier directions are
orthogonal to $\mathbf G(\mathbf x)^\top$, and the objective is separable
in $(r,s)$, the CBF quadratic-program constraint is strictly inactive in a neighborhood
of $s=0$. Consequently, its gradient is uniformly nonzero whenever the
constraint is binding. Moreover,
$\mathbf G(\mathbf x)\mathbf H(\mathbf x)^\top h(\mathbf x)=0$ on the
boundary of $\Sg^\star$. The mixed MFCQ direction can be chosen as
$\operatorname{col}(-1,-1)$ at $s=1$ and
$\operatorname{col}(1,1)$ at $s=-1$.

Since $r$ and $s$ are linear in $\mathbf x$, the Lagrangian Hessian along
the quadratic-program solution satisfies
\(
\begin{aligned}
\mathbf Q(\mathbf x,z)
\succ0.
\end{aligned}
\)
Thus, all the assumptions of
Theorem~\ref{thm:safe_component_convergence} hold on $\Sg^\star$.

We select $K_p=K_i=1$, $\mathbf z(0)=\mathbf0$, and three different
initial conditions. As shown in
Fig.~\ref{fig:two_dimensional_examples}(b), the trajectories start from
widely separated points in the noncompact set $\Sg^\star$, remain inside
$\Sg^\star$, and converge to $\mathbf x^\star$. These results illustrate
the safe-component convergence established in
Theorem~\ref{thm:safe_component_convergence}.

\subsection{Safe Platoon Coordination and Comparative Evaluation}
\label{subsec:platoon_simulation}

We further consider an application-oriented example involving a platoon of
$N=6$ vehicles \cite{Wang2016MultiObjective}. Each vehicle is modeled by the
single-integrator dynamics $\dot p_i=v_i$, $i=1,\ldots,N$, where
$p_i\in\mathbb R$ denotes its longitudinal position. Define
$\mathbf p=\operatorname{col}(p_1,\ldots,p_N)$ and the intervehicle distances
$q_i=p_{i+1}-p_i$, $i=1,\ldots,N-1$.
Let $\mathbf B\in\mathbb R^{(N-1)\times N}$ denote the incidence matrix of
the path graph, whose $i$th row is
$\mathbf b_i^\top=\mathbf e_{i+1}^\top-\mathbf e_i^\top$. Then
$\mathbf q=\mathbf B\mathbf p$.

The desired vehicle positions are selected as
$p_i^{\mathrm d}=3\left(i-\frac{N+1}{2}\right)$, $i=1,\ldots,N$, and the objective
function is $f(\mathbf p)=\frac{1}{2}\|\mathbf p-\mathbf p^{\mathrm d}\|^2$.
The centroid condition is imposed as the equality constraint
$h(\mathbf p)=\frac{1}{N}\mathbf 1_N^\top\mathbf p=0$.
This constraint specifies the final platoon location but is allowed to be
violated during the transient.

The inequality constraints are defined by
\[
\begin{aligned}
g_i^-(\mathbf p)
&=d_{\min}^2-q_i^2\leq0,\\
g_i^+(\mathbf p)
&=q_i^2-d_{\max}^2\leq0,
\qquad i=1,\ldots,N-1,
\end{aligned}
\]
where $d_{\min}=1$ and $d_{\max}=2$. The lower constraints prevent
collisions, whereas the upper constraints preserve the communication
links between neighboring vehicles.

Because squared-distance constraints permit both positive and negative
relative positions, we consider the ordered connected component
\(
\Sg^\star
=
\left\{
\mathbf p\in\mathbb R^N:
1\leq q_i\leq2,\quad i=1,\ldots,N-1
\right\}.
\)
The set $\Sg^\star$ is closed and connected, and
$\mathcal F^\star=\{\mathbf p\in\Sg^\star:
\mathbf 1_N^\top\mathbf p=0\}$ is compact. The unique optimizer in this
component is $p_i^\star=2\left(i-\frac{N+1}{2}\right)$,
$i=1,\ldots,N$,
at which all upper-distance constraints are active.

The structure of this example also allows the assumptions of
Theorem~\ref{thm:safe_component_convergence} to be verified directly.
For all simulations, we use $K_p=K_i=N$.

\subsubsection{Comparison with the A-ICBF-OR flow for different $\tau$}
\label{subsubsec:tau_comparison}
We first compare the exact ICBF-OR flow with the A-ICBF-OR
flow for $\tau\in\{0.05,\,0.5,\,2\}$. The exact
ICBF-OR trajectory is represented by $\tau=0$, corresponding to the algebraic
fast-solver limit discussed in
Subsection~\ref{subsec:relations_solution_methods}.

\begin{figure}[t]
\centering
\includegraphics[width=\linewidth]{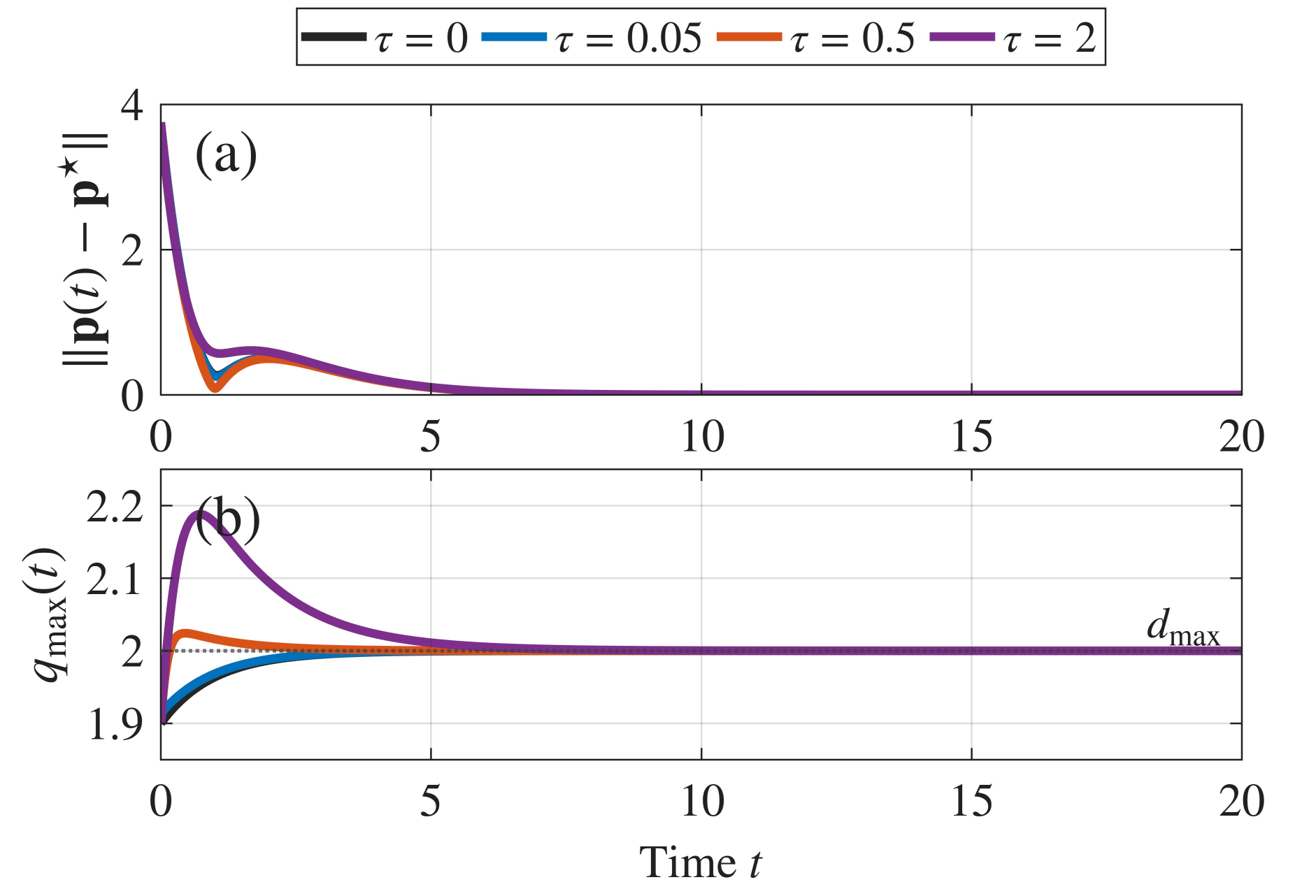}
\caption{Platoon coordination using the  ICBF-OR flow
and the A-ICBF-OR flow with different values of $\tau$.}
\label{fig:platoon_tau}
\end{figure}

Fig.~\ref{fig:platoon_tau}(a) reports the state error
$\|\mathbf p(t)-\mathbf p^\star\|$, while
Fig.~\ref{fig:platoon_tau}(b) shows the largest intervehicle distance,
$q_{\max}(t)=\max_{i=1,\ldots,N-1}q_i(t)$, which characterizes the
violation of the upper-distance constraints. All trajectories converge
to the optimizer $\mathbf p^\star$, illustrating the convergence result
in the proposed theorems. As $\tau$ decreases,
the A-ICBF-OR trajectories approach the exact ICBF-OR
trajectory, and the peak constraint violation decreases accordingly.
For larger values of $\tau$, the multiplier tracking error causes
$q_{\max}(t)$ to temporarily exceed $d_{\max}$. In the present
simulation, choosing $\tau=0.05$ eliminates the observed constraint
violation and closely reproduces the exact ICBF-OR trajectory. These
results are consistent with the $O(\tau)$ violation bound established
in Proposition~\ref{prop:dynamic_constraint_violation}.

\begin{figure}[t]
    \centering
    \includegraphics[width=\linewidth]
    {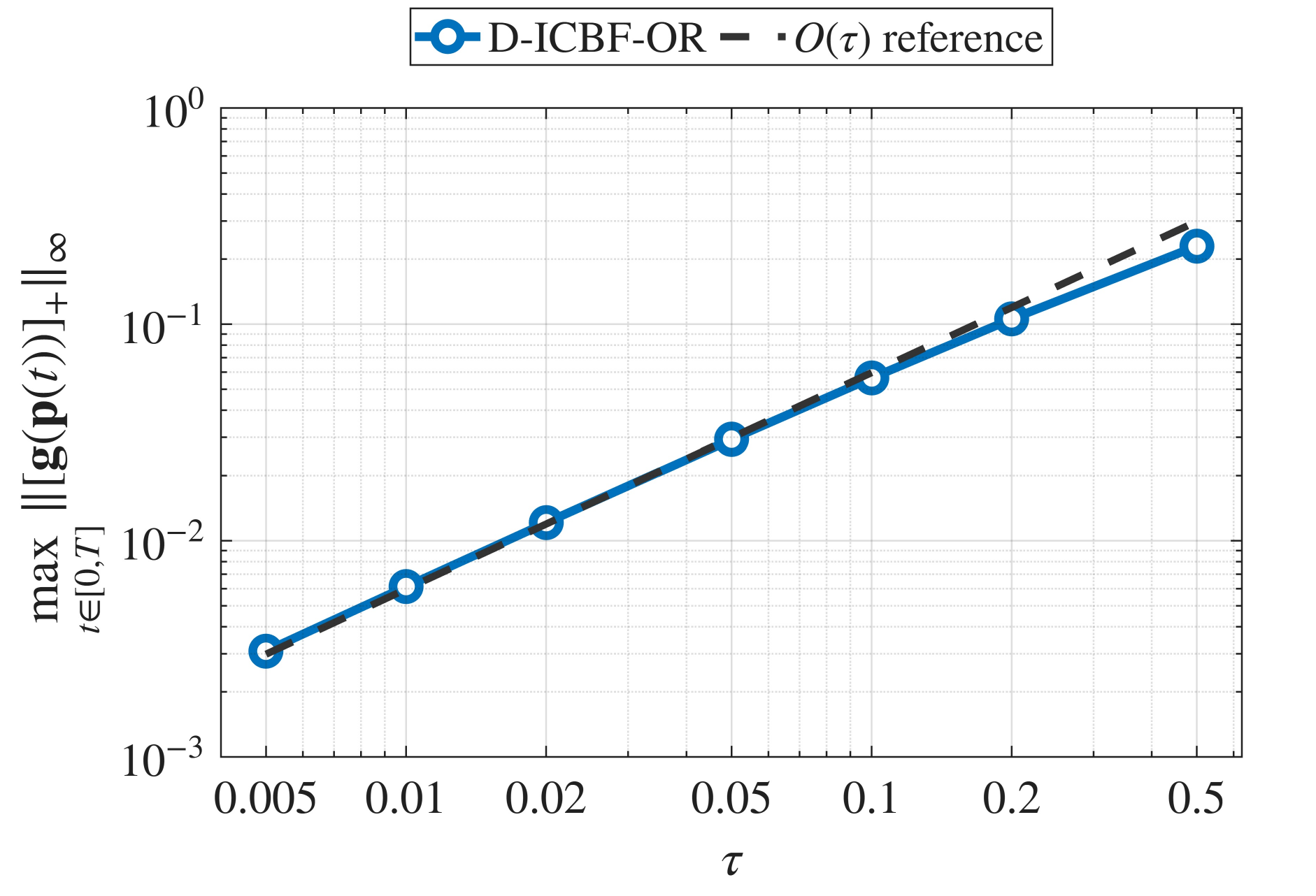}
    \caption{Maximum constraint violation of A-ICBF-OR for different values of $\tau$.}
    \label{fig:tau_violation}
\end{figure}

To further examine the $O(\tau)$ constraint-violation bound established in Proposition \ref{prop:dynamic_constraint_violation}, Fig.~\ref{fig:tau_violation} plots $\max_{t\in[0,T]}\|[\mathbf g(\mathbf p(t))]_+\|_\infty$ against $\tau$. To make the $\tau$-dependent transient violation observable, we set $q_1(0)=q_5(0)=d_{\max}$ only for this test, while keeping all other settings unchanged. It can be observed that the maximum violation decreases approximately linearly with $\tau$, verifying Proposition \ref{prop:dynamic_constraint_violation}. 

\subsubsection{Comparison with existing flows}
\label{subsubsec:flow_comparison}
We next compare the proposed ICBF-OR flow and its approximate implementation,
A-ICBF-OR with $\tau=0.05$, with the following methods:
\begin{itemize}
\item the P-PD flow of
\cite{Tang2020SemiGlobal};
\item the safe gradient flow of
\cite{allibhoy2024control}, denoted by SGF.
\end{itemize}
The same initial primal state and inequality CBF gain are used for all
methods. For A-ICBF-OR, the constraint-tightening strategy introduced in
Section~\ref{sec:cbf_qp_solution} is employed to ensure safety of
the inequality constraints throughout the transient.

\begin{figure}[t]
\centering
\includegraphics[width=\linewidth]
{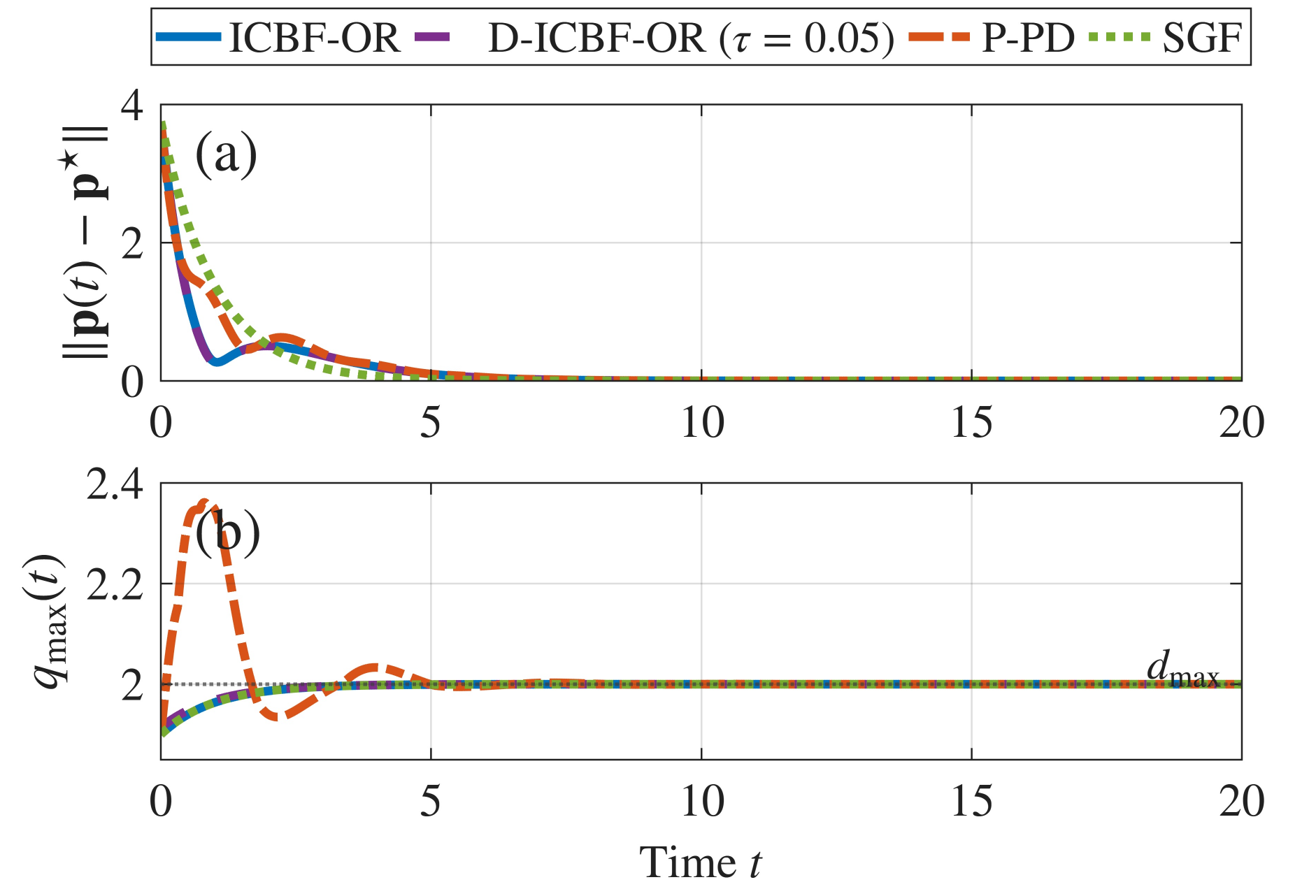}
\caption{Platoon coordination using ICBF-OR, A-ICBF-OR, P-PD, and SGF.}
\label{fig:platoon_comparison}
\end{figure}

Fig.~\ref{fig:platoon_comparison} reports the same performance
quantities as Fig.~\ref{fig:platoon_tau}. All four methods converge to
the same optimizer, with the common initial state chosen sufficiently close
to it. However, P-PD fails to preserve inequality feasibility, as
$q_{\max}(t)$ temporarily exceeds the prescribed upper bound
$d_{\max}$. By contrast, ICBF-OR, A-ICBF-OR with $\tau=0.05$, and SGF maintain the
inequality constraints throughout this simulation. SGF also
incorporates the equality conditions into its safety filter,
whereas the proposed flows regulate the equality residuals
asymptotically. Under the conditions stated in
Theorem~\ref{thm:safe_component_convergence}, ICBF-OR additionally has a
convergence guarantee from the considered connected component
of the inequality-feasible set.

Table~\ref{tab:computation_time} reports the average computation time per vector-field evaluation, where ICBF-OR is implemented using the analytical and iterative solution methods described in Section~\ref{sec:algorithm}. The analytical implementation of ICBF-OR achieves the fastest computation. Together with its ability to guarantee inequality safety, this makes it the preferred choice for practical implementation whenever an analytical solution is available. A-ICBF-OR has the second-lowest computational cost, comparable to that of P-PD and substantially lower than that of iterative ICBF-OR. Therefore, when an analytical solution of the ICBF-OR quadratic program is unavailable, A-ICBF-OR with a sufficiently small $\tau$ provides an attractive alternative for maintaining the prescribed inequality-safety tolerance with low computational cost. In contrast, the iterative implementation of SGF is slower than all the other methods in this example. These measurements
illustrate an implementation-dependent computational advantage of the proposed flows. 
\begin{table}[t]
    \centering
    \caption{Average computation time per vector-field evaluation.}
    \label{tab:computation_time}
    \footnotesize
    \setlength{\tabcolsep}{5pt}
    \begin{tabular}{lcc}
        \hline
        Method & Implementation & Time (ms) \\
        \hline
        ICBF-OR   & Analytical & $0.051$ \\
        ICBF-OR   & Iterative  & $0.775$ \\
        A-ICBF-OR & $\tau=0.05$ & $0.069$ \\
        P-PD      & --          & $0.092$ \\
        SGF       & Iterative           & $2.240$ \\
        \hline
    \end{tabular}
\end{table}

\subsection{Stability-Constrained Imitation Learning for Servo Control}
\label{subsec:servo_learning}

We finally consider a stability-constrained imitation-learning problem
for servo-controller design \cite{Havens2021Imitation,Tu2022Sample}. Consider the double-integrator system
\(\ddot q=u\)
with the affine proportional--derivative controller
\(
u=-k_{\rm p}q-k_{\rm d}\dot q+c,
\)
where
$\boldsymbol{\theta}
=\operatorname{col}(k_{\rm p},k_{\rm d},c)\in\mathbb R^3$
collects the controller parameters to be learned.

Given a demonstration dataset
$\mathcal D=\{(q_\ell,\dot q_\ell,u_\ell^{\rm E})\}_{\ell=1}^{M}$,
the controller is learned by minimizing the empirical imitation loss
\[
f(\boldsymbol{\theta})
=
\frac{1}{2M}
\sum_{\ell=1}^{M}
\left(
u_\ell^{\rm E}
+k_{\rm p}q_\ell
+k_{\rm d}\dot q_\ell-c
\right)^2.
\]

Let $q_{\rm r}$ denote the desired reference position. For $q_{\rm r}$
to be an equilibrium of the learned closed-loop system, the controller
parameters must satisfy
\[
h(\boldsymbol{\theta})
=
c-k_{\rm p}q_{\rm r}
=0.
\]
Moreover, the controller is required to remain stabilizing throughout
the learning process, so that the system remains stable even if learning
is terminated at any intermediate time. The characteristic polynomial of the closed-loop
system is
\(
s^2+k_{\rm d}s+k_{\rm p}.
\)
Requiring its roots to lie in
$\{s\in\mathbb C:\operatorname{Re}(s)\leq-\delta\}$, where
$\delta>0$ is the prescribed stability margin, gives
\begin{equation}
\begin{aligned}
g_1(\boldsymbol{\theta})
&=2\delta-k_{\rm d}\leq0,\\
g_2(\boldsymbol{\theta})
&=\delta k_{\rm d}-\delta^2-k_{\rm p}\leq0.
\end{aligned}
\label{eq:servo_stability_constraints}
\end{equation}

For the simulation, we select \(
q_{\rm r}=1,\
\delta=0.5.\)
The resulting constrained learning problem has the unique optimizer
\(
\boldsymbol{\theta}^\star
=
\operatorname{col}(0.25,1,0.25).\)
At this point, both stability constraints in
\eqref{eq:servo_stability_constraints} are active. A corresponding KKT
triple is obtained with $z^\star=-0.15$ and
$\boldsymbol{\mu}^\star=\operatorname{col}(0.65,0.30)$.
Moreover, the equality Jacobian has constant rank and the two active
inequality gradients are linearly independent. The demonstration data are normalized such that the empirical feature
Gram matrix is the identity matrix. Consequently,
$\nabla^2 f(\boldsymbol{\theta})=\mathbf I_3$. Therefore, the local
regularity and curvature conditions required by the proposed flows hold.

We use $K_p=K_i=1$ and select
\(
\boldsymbol{\theta}(0)
=
\operatorname{col}(0.25,1,0.45),
\
z(0)=0,
\
\boldsymbol{\lambda}(0)=\mathbf0.
\)
The exact ICBF-OR flow is 
represented by $\tau=0$, while the A-ICBF-OR flow is simulated with
$\tau\in\{0.05,0.5,2\}$.

\begin{figure}[t]
\centering
\includegraphics[width=\linewidth]
{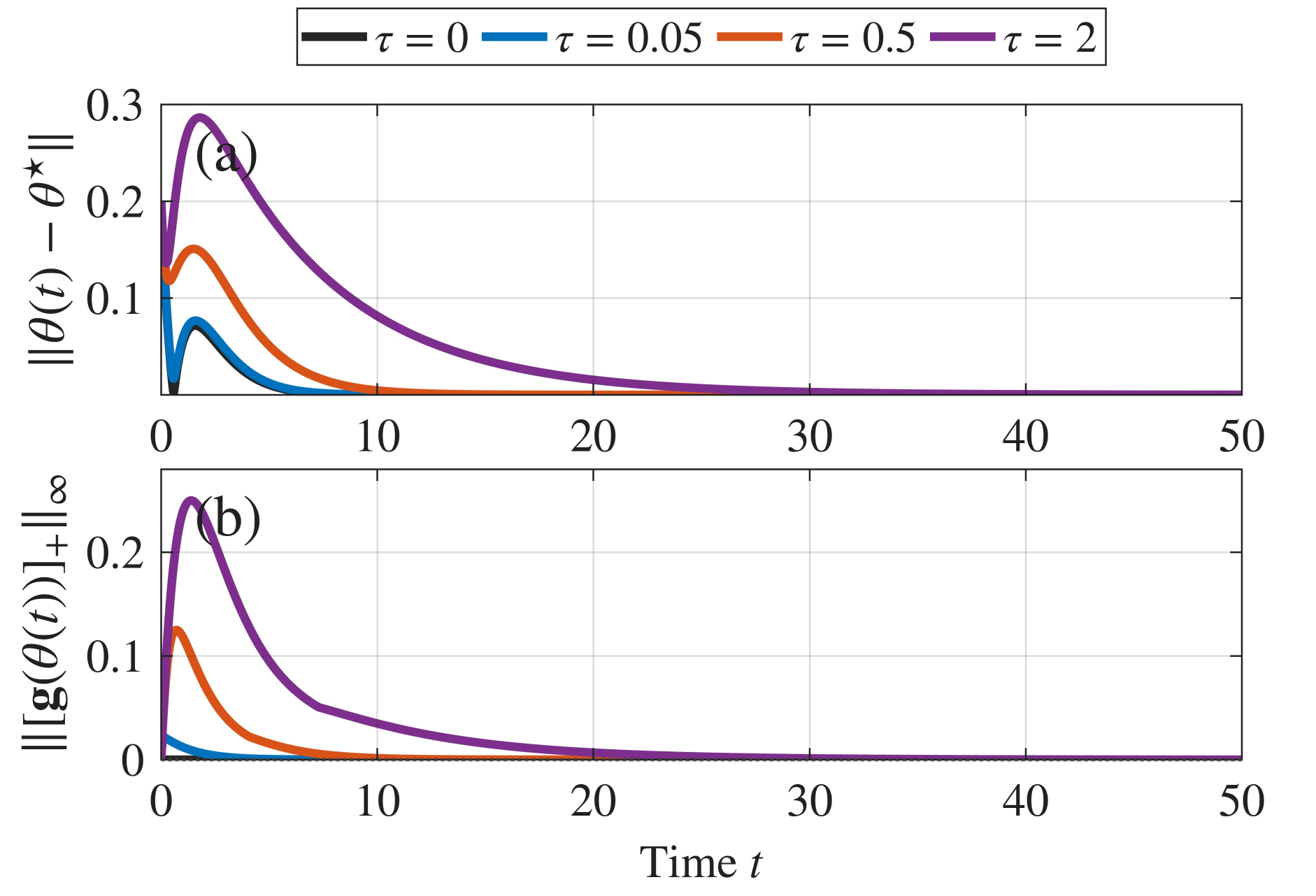}
\caption{Stability-constrained imitation learning for servo control using
ICBF-OR and A-ICBF-OR with different values of $\tau$.}
\label{fig:servo_learning_tau}
\end{figure}

Fig.~\ref{fig:servo_learning_tau}(a) shows the parameter-learning error
$\|\boldsymbol{\theta}(t)-\boldsymbol{\theta}^\star\|$, while
Fig.~\ref{fig:servo_learning_tau}(b) reports the maximum instantaneous
stability-constraint violation
$\|[\mathbf g(\boldsymbol{\theta}(t))]_+\|_\infty$.
All the tested flows converge to the optimal controller parameters, verifying the convergence results
in the proposed theorems.
As $\tau$ decreases, the A-ICBF-OR trajectory approaches the exact ICBF-OR trajectory, which verifies the relation proposed in Section \ref{subsec:relations_solution_methods}. The ICBF-OR flow preserves the prescribed
stability region throughout learning, whereas A-ICBF-OR may exhibit a
transient violation. The peak
violation decreases as $\tau$ becomes smaller, consistently with
Proposition~\ref{prop:dynamic_constraint_violation}.

\section{Conclusion}
\label{sec:conclusion}

In this paper, we have proposed the ICBF-OR flow, which combines proportional--integral output regulation of the equality residuals with an inequality-only CBF safety filter. We have established its safety and convergence properties without requiring a combined LICQ. We have also developed the A-ICBF-OR implementation and have provided a unified formulation connecting ICBF-OR, A-ICBF-OR, and projected primal--dual dynamics.

The present framework also opens several avenues for further research.
An important direction for future research is to develop a
constraint-selective framework that assigns enforcement mechanisms
according to the practical roles of individual constraints. Another direction is to incorporate advanced CBF variants developed for different control objectives \cite{Taylor2020Adaptive,Dawson2023Safe}, such as robustness and high-relative-degree constraints, and to investigate how their properties can benefit constrained optimization flows. Finally, extending the proposed framework to representative structured constrained optimization problems, particularly distributed optimization over networks, will further clarify its applicability and scalability.

\appendix

\section{Proof of Proposition \ref{prop:qp_feasibility}}

For any \(\mathbf x\in\Sg\), the candidate \(\mathbf v=\mathbf0\) satisfies
$\mathbf G(\mathbf x)\mathbf0-
\boldsymbol{\alpha}(-\mathbf g(\mathbf x))
=-\boldsymbol{\alpha}(-\mathbf g(\mathbf x))\leq\mathbf0$,
where the inequality follows from $-\mathbf g(\mathbf x)\geq\mathbf0$ and
the definition of $\boldsymbol{\alpha}$.
Hence, the feasible set is nonempty, closed, and convex. Given that the
objective in \eqref{eq:cbf_qp} is coercive and strongly convex in \(\mathbf v\), a unique optimizer exists.

\section{Proof of Proposition \ref{prop:forward_invariance}}

By Proposition~\ref{prop:qp_feasibility}, feasibility of
$\mathbf v^\star$ in \eqref{eq:cbf_qp} and the chain rule give, along
every solution, $\dot g_j=\nabla g_j(\mathbf x)^\top\mathbf v^\star
\leq\alpha_j(-g_j(\mathbf x))$.
Setting $b_j(t):=-g_j(\mathbf x(t))$ gives
$\dot b_j(t)\geq-\alpha_j(b_j(t))$. By the comparison lemma and the
definition of $\eta_j$, $b_j(t)\geq\eta_j(t;b_j(0))$.
Since $b_j(0)\geq0$ and the nonnegative half-line is invariant under
$\dot\eta_j=-\alpha_j(\eta_j)$, one has
$\eta_j(t;b_j(0))\geq0$. Recalling that $b_j=-g_j$ yields
\eqref{eq:constraint_comparison}.

\section{Proof of Lemma~\ref{lem:equilibrium_KKT}}
{
Suppose first that \((\mathbf x^\star,\mathbf z^\star)\) is an equilibrium. Since
\(\mathbf K_i\succ0\), the equality \(\dot{\mathbf z}=\mathbf0\) implies
\(\mathbf h(\mathbf x^\star)=\mathbf0\). Moreover,
\(\mathbf v^\star(\mathbf x^\star,\mathbf z^\star)=\mathbf0\). The feasible set of
\eqref{eq:cbf_qp} is polyhedral in \(\mathbf v\). Hence, the first-order
optimality condition at \(\mathbf v=\mathbf0\) yields a multiplier
\(\boldsymbol{\mu}^\star\geq\mathbf0\) satisfying
$-\vpd(\mathbf x^\star,\mathbf z^\star)
+\mathbf G(\mathbf x^\star)^\top\boldsymbol{\mu}^\star=\mathbf0$ and
$\mu_j^\star\alpha_j(-g_j(\mathbf x^\star))=0$ for every $j$.
Using $\mathbf h(\mathbf x^\star)=\mathbf0$ in \eqref{eq:vpd},
the first equality becomes
$\nabla f(\mathbf x^\star)+\mathbf H(\mathbf x^\star)^\top\mathbf z^\star
+\mathbf G(\mathbf x^\star)^\top\boldsymbol{\mu}^\star=\mathbf0$,
which proves \eqref{eq:KKTa}.
Because $\mathbf x^\star\in\Sg$ and $\alpha_j$ is strictly increasing with
$\alpha_j(0)=0$, $\alpha_j(-g_j(\mathbf x^\star))=0$ if and only if
$g_j(\mathbf x^\star)=0$.
Hence, the second equality implies
$\mu_j^\star g_j(\mathbf x^\star)=0$ for every $j$. Together with \(\boldsymbol{\mu}^\star\geq\mathbf0\), \eqref{eq:KKTb} and \eqref{eq:KKTc} are satisfied. Thus,
$(\mathbf x^\star,\mathbf z^\star,\boldsymbol{\mu}^\star)$ is a KKT triple.

Conversely, let \((\mathbf x^\star,\mathbf z^\star,\boldsymbol{\mu}^\star)\) be a KKT triple.
Because \(\mathbf h(\mathbf x^\star)=\mathbf0\), KKT stationarity gives
$\vpd(\mathbf x^\star,\mathbf z^\star)
=\mathbf G(\mathbf x^\star)^\top\boldsymbol{\mu}^\star$.
Moreover, the original complementarity condition and
$\mathbf x^\star\in\Sg$ yield
$\mu_j^\star\alpha_j(-g_j(\mathbf x^\star))=0$ for every $j$.
The pair \((\mathbf v,\boldsymbol{\mu})=(\mathbf0,\boldsymbol{\mu}^\star)\) therefore satisfies the KKT
conditions of \eqref{eq:cbf_qp}. Since the objective of \eqref{eq:cbf_qp} is
strongly convex, \(\mathbf v^\star(\mathbf x^\star,\mathbf z^\star)=\mathbf0\) is its unique
optimizer. In addition, \(\dot{\mathbf z}=\mathbf K_i\mathbf h(\mathbf x^\star)=\mathbf0\), proving
that \((\mathbf x^\star,\mathbf z^\star)\) is an equilibrium.
}

\section{Proof of Lemma~\ref{lem:local_kkt_set}}
\label{app:proof_local_kkt}

\noindent\emph{Sufficiency.}
Since $\mathbf N^\star$ has orthonormal columns spanning
$\ker\mathbf A^\star$, the condition
$\mathbf B^{\star\top}\boldsymbol{\eta}=\mathbf0$ implies
\(
\mathbf N^{\star\top}\mathbf H^{\star\top}\boldsymbol{\eta}
=
\mathbf B^{\star\top}\boldsymbol{\eta}
=\mathbf0.
\)
Therefore,
\(
\mathbf H^{\star\top}\boldsymbol{\eta}
\in
\bigl(\ker\mathbf A^\star\bigr)^\perp
=
\operatorname{range}(\mathbf A^{\star\top}).
\)
Moreover, Assumption~\ref{ass:B3} gives
{
\(
\mathbf I-
\mathbf A^{\star\top}
\left(
\mathbf A^\star\mathbf A^{\star\top}
\right)^{-1}
\mathbf A^\star
=
\mathbf N^\star\mathbf N^{\star\top}.
\)
}
Using \eqref{eq:local_inequality_multiplier} and the stationarity of the
reference KKT triple yields
{
\begin{align*}
&\nabla f(\mathbf x^\star)
+\mathbf H^{\star\top}
\left(\mathbf z^\star+\boldsymbol{\eta}\right)
+\mathbf A^{\star\top}
\boldsymbol{\mu}_{\mathcal A^\star}(\boldsymbol{\eta})
\\
=&
\nabla f(\mathbf x^\star)
+\mathbf H^{\star\top}\mathbf z^\star
+\mathbf A^{\star\top}
\boldsymbol{\mu}_{\mathcal A^\star}^\star
\\&+
\left[
\mathbf I-
\mathbf A^{\star\top}
\left(
\mathbf A^\star\mathbf A^{\star\top}
\right)^{-1}
\mathbf A^\star
\right]
\mathbf H^{\star\top}\boldsymbol{\eta}
\\
=&
\nabla f(\mathbf x^\star)
+\mathbf H^{\star\top}\mathbf z^\star
+\mathbf A^{\star\top}
\boldsymbol{\mu}_{\mathcal A^\star}^\star
+\mathbf N^\star\mathbf B^{\star\top}\boldsymbol{\eta}
=\mathbf0.
\end{align*}
}
Thus, \eqref{eq:KKTa} holds.

Since $\mathbf x^\star\in\mathcal F$, primal feasibility holds. By the
definition of $\mathcal M^\star$ and
\eqref{eq:local_inequality_multiplier},
$\boldsymbol{\mu}(\boldsymbol{\eta})\geq\mathbf0$, so \eqref{eq:KKTb}
holds. Finally, $g_j(\mathbf x^\star)=0$ for every
$j\in\mathcal A^\star$, whereas $\mu_j(\boldsymbol{\eta})=0$ for every
$j\notin\mathcal A^\star$. Thus, \eqref{eq:KKTc} also holds, and
$(\mathbf x^\star,\mathbf z^\star+\boldsymbol{\eta},
\boldsymbol{\mu}(\boldsymbol{\eta}))$ is a KKT triple.

\medskip
\noindent\emph{Necessity.} Let $(\mathbf x^\star,\mathbf z,\boldsymbol{\mu})$ be a
KKT triple and set $\boldsymbol{\eta}:=\mathbf z-\mathbf z^\star$. For
every $j\notin\mathcal A^\star$, $g_j(\mathbf x^\star)<0$ and
\eqref{eq:KKTc} give $\mu_j=\mu_j^\star=0$. Subtracting \eqref{eq:KKTa}
evaluated at $(\mathbf x^\star,\mathbf z^\star,\boldsymbol{\mu}^\star)$
from \eqref{eq:KKTa} evaluated at
$(\mathbf x^\star,\mathbf z,\boldsymbol{\mu})$ yields
\(
\mathbf H^{\star\top}\boldsymbol{\eta}
+\mathbf A^{\star\top}
\bigl(
\boldsymbol{\mu}_{\mathcal A^\star}
-\boldsymbol{\mu}_{\mathcal A^\star}^\star
\bigr)
=\mathbf0.
\)
Premultiplying by $\mathbf N^{\star\top}$ and using
$\mathbf A^\star\mathbf N^\star=\mathbf0$ gives
$\mathbf B^{\star\top}\boldsymbol{\eta}=\mathbf0$. Premultiplying by
$(\mathbf A^\star\mathbf A^{\star\top})^{-1}\mathbf A^\star$, which is
well defined by Assumption~\ref{ass:B3}, gives
$\boldsymbol{\mu}_{\mathcal A^\star}
=\boldsymbol{\mu}_{\mathcal A^\star}(\boldsymbol{\eta})$. Hence,
$\boldsymbol{\mu}=\boldsymbol{\mu}(\boldsymbol{\eta})$, and
$\boldsymbol{\mu}\geq\mathbf0$ implies
$\boldsymbol{\eta}\in\mathcal M^\star$.

\section{Proof of Theorem~\ref{thm:local_convergence}}
\label{app:proof_local_convergence}

\noindent
\emph{Step 1: Local uniqueness and smoothness of the CBF quadratic-program solution.}

We first establish that, in a neighborhood of the reference KKT triple,
the solution mapping of the CBF quadratic program
\eqref{eq:cbf_qp},
\(
(\mathbf x,\mathbf z)
\mapsto
(\mathbf v^\star(\mathbf x,\mathbf z),
\boldsymbol{\lambda}^\star(\mathbf x,\mathbf z)),
\)
is single-valued and continuously differentiable, with its binding set
remaining \(\mathcal A^\star\).
For fixed \((\mathbf x,\mathbf z)\), let
\(\boldsymbol{\lambda}\in\R_{\geq0}^m\) denote the multiplier associated
with the constraints of \eqref{eq:cbf_qp}. The KKT conditions of this
quadratic program are
\begin{subequations}
\label{eq:cbf_qp_KKT}
\begin{align}
\mathbf v-\mathbf v_{\mathrm{pd}}(\mathbf x,\mathbf z)
+\mathbf G(\mathbf x)^\top\boldsymbol{\lambda}
&=\mathbf0,
\label{eq:cbf_qp_KKTa}\\
\mathbf G(\mathbf x)\mathbf v
-\boldsymbol{\alpha}\bigl(-\mathbf g(\mathbf x)\bigr)
&\leq\mathbf0,
\label{eq:cbf_qp_KKTb}\\
\boldsymbol{\lambda}
&\geq\mathbf0,
\label{eq:cbf_qp_KKTc}\\
\lambda_j
\left[
\nabla g_j(\mathbf x)^\top\mathbf v
-\alpha_j\bigl(-g_j(\mathbf x)\bigr)
\right]
&=0,
\quad j=1,\ldots,m.
\label{eq:cbf_qp_KKTd}
\end{align}
\end{subequations}

We first verify \eqref{eq:cbf_qp_KKT} at the reference KKT triple
\((\mathbf x^\star,\mathbf z^\star,\boldsymbol{\mu}^\star)\). Since
\(\mathbf h(\mathbf x^\star)=\mathbf0\), it follows from
\eqref{eq:vpd} and \eqref{eq:KKTa} that
\begin{align}
\mathbf v_{\mathrm{pd}}(\mathbf x^\star,\mathbf z^\star)
&=-\nabla f(\mathbf x^\star)
-\mathbf H^{\star\top}\mathbf z^\star
\nonumber\\
&=\mathbf G(\mathbf x^\star)^\top\boldsymbol{\mu}^\star
=\mathbf A^{\star\top}
\boldsymbol{\mu}_{\mathcal A^\star}^\star,
\label{eq:appendix_vpd_at_kkt}
\end{align}
where
\(\boldsymbol{\mu}_{(\mathcal A^\star)^c}^\star=\mathbf0\) follows
from \eqref{eq:KKTc}.

Set \(\mathbf v=\mathbf0\) and
\(\boldsymbol{\lambda}=\boldsymbol{\mu}^\star\). Then
\eqref{eq:appendix_vpd_at_kkt} verifies
\eqref{eq:cbf_qp_KKTa}. For every \(j\in\mathcal A^\star\), one has
\(g_j(\mathbf x^\star)=0\), and hence the corresponding constraint
residual in \eqref{eq:cbf_qp_KKTb} is zero. {For every
\(j\notin\mathcal A^\star\), one has \(g_j(\mathbf x^\star)<0\), and
the extended class-\(\mathcal K\) property gives
$\nabla g_j(\mathbf x^\star)^\top\mathbf v
-\alpha_j(-g_j(\mathbf x^\star))
=-\alpha_j(-g_j(\mathbf x^\star))<0$.}
Thus, \eqref{eq:cbf_qp_KKTb} holds. Conditions
\eqref{eq:cbf_qp_KKTc} and \eqref{eq:cbf_qp_KKTd} follow from
\eqref{eq:KKTb}--\eqref{eq:KKTc}. Therefore,
\begin{equation}
\label{eq:reference_qp_pair}
\left(
\mathbf v^\star(\mathbf x^\star,\mathbf z^\star),
\boldsymbol{\lambda}^\star
\right)
=
\left(\mathbf0,\boldsymbol{\mu}^\star\right)
\end{equation}
is a solution of \eqref{eq:cbf_qp} at
\((\mathbf x^\star,\mathbf z^\star)\).

The optimizer \(\mathbf v^\star(\mathbf x^\star,\mathbf z^\star)\)
is unique because the objective of \eqref{eq:cbf_qp} is strongly convex
in \(\mathbf v\). The associated multiplier
\(\boldsymbol{\lambda}^\star\) is also unique. Indeed, the constraints
indexed by \((\mathcal A^\star)^c\) are strict at
\eqref{eq:reference_qp_pair}, so \eqref{eq:cbf_qp_KKTd} implies
\(\boldsymbol{\lambda}_{(\mathcal A^\star)^c}^\star=\mathbf0\).
{If \(\widetilde{\boldsymbol{\lambda}}^\star\) is another quadratic-program multiplier,
then \eqref{eq:cbf_qp_KKTa} gives
$\mathbf A^{\star\top}(
\widetilde{\boldsymbol{\lambda}}_{\mathcal A^\star}^\star
-\boldsymbol{\lambda}_{\mathcal A^\star}^\star
)=\mathbf0$.}
Assumption~\ref{ass:B3} implies that
\(\mathbf A^{\star\top}\) has full column rank. Hence,
\(\widetilde{\boldsymbol{\lambda}}_{\mathcal A^\star}^\star
=\boldsymbol{\lambda}_{\mathcal A^\star}^\star\), which proves the
uniqueness of \(\boldsymbol{\lambda}^\star\).

The Jacobian of \eqref{eq:cbf_qp_KKTa} and the active constraints in
\eqref{eq:cbf_qp_KKTb}, with respect to
\((\mathbf v,\boldsymbol{\lambda}_{\mathcal A^\star})\) and evaluated
at \eqref{eq:reference_qp_pair}, is
\begin{equation}
\label{eq:appendix_qp_kkt_matrix}
\begin{bmatrix}
\mathbf I&\mathbf A^{\star\top}\\
\mathbf A^\star&\mathbf0
\end{bmatrix}.
\end{equation}
By Assumption~\ref{ass:B3}, the matrix in
\eqref{eq:appendix_qp_kkt_matrix} is nonsingular. The implicit function
theorem therefore yields unique continuously differentiable local
mappings satisfying these active KKT equations,
\begin{equation}
\label{eq:multi_map}
(\mathbf x,\mathbf z)
\mapsto
\left(
\mathbf v^\star(\mathbf x,\mathbf z),
\boldsymbol{\lambda}_{\mathcal A^\star}^\star(\mathbf x,\mathbf z)
\right),
\end{equation}
around \((\mathbf x^\star,\mathbf z^\star)\).

It remains to verify that the quadratic-program binding set is unchanged locally. By
Assumption~\ref{ass:B4} and \eqref{eq:reference_qp_pair},
\(\boldsymbol{\lambda}_{\mathcal A^\star}^\star
=\boldsymbol{\mu}_{\mathcal A^\star}^\star>\mathbf0\).
Continuity of \eqref{eq:multi_map} ensures that these multipliers remain
positive in a sufficiently small neighborhood. By
\eqref{eq:cbf_qp_KKTd}, the corresponding constraints remain active.
Moreover, the constraints indexed by \((\mathcal A^\star)^c\), which
are strict at the reference point, remain inactive by continuity.
Hence, the binding set of \eqref{eq:cbf_qp} remains exactly
\(\mathcal A^\star\), and \eqref{eq:multi_map}, completed by
\(\boldsymbol{\lambda}_{(\mathcal A^\star)^c}^\star=\mathbf0\), is
the unique local solution mapping of \eqref{eq:cbf_qp}.

\medskip

\noindent
\emph{Step 2: Linearization of the dynamics.}

By Step~1, the local solution mapping
\eqref{eq:multi_map} is continuously differentiable and the quadratic-program active
set remains \(\mathcal A^\star\). Let
\(\delta\mathbf x:=\mathbf x-\mathbf x^\star\) and
\(\delta\mathbf z:=\mathbf z-\mathbf z^\star\), and let
\(\delta\mathbf v\) and
\(\delta\boldsymbol{\lambda}_{\mathcal A^\star}\) denote the
corresponding variations of the quadratic-program primal and active dual variables.

Linearizing \eqref{eq:cbf_qp_KKTa} and the active constraints in
\eqref{eq:cbf_qp_KKTb} around \eqref{eq:reference_qp_pair} gives
\begin{equation}
\begin{aligned}
\delta\mathbf v
+\left(
\mathbf Q^\star
+\mathbf H^{\star\top}\mathbf K_p\mathbf H^\star
\right)\delta\mathbf x
+\mathbf H^{\star\top}\delta\mathbf z
+\mathbf A^{\star\top}
\delta\boldsymbol{\lambda}_{\mathcal A^\star}
&=\mathbf0,\\
\mathbf A^\star\delta\mathbf v
+\boldsymbol{\Gamma}_{\mathcal A^\star}
\mathbf A^\star\delta\mathbf x
&=\mathbf0,
\label{eq:appendix_linearized_constraint}
\end{aligned}
\end{equation}
where
\(\boldsymbol{\Gamma}_{\mathcal A^\star}
:=\operatorname{diag}(\alpha_j'(0):j\in\mathcal A^\star)
\succ\mathbf0\) by the standing condition on $\alpha_j$. The first 
equation follows from
\eqref{eq:vpd} and the definitions of $\mathbf H^\star$ and
$\mathbf Q^\star$, and the second equation follows by differentiating
the active constraints at \eqref{eq:reference_qp_pair}.

{By Assumption~\ref{ass:B3}, the right inverse
$\mathbf A^{\star\dagger}:=\mathbf A^{\star\top}
(\mathbf A^\star\mathbf A^{\star\top})^{-1}$ is well defined and satisfies
\(\mathbf I-\mathbf A^{\star\dagger}\mathbf A^\star
=\mathbf N^\star\mathbf N^{\star\top}\).} Decomposing
\(\delta\mathbf v\) according to these complementary projections and
using
\eqref{eq:appendix_linearized_constraint} gives
\begin{equation}
\label{eq:appendix_linearized_flow}
\begin{aligned}
\delta\dot{\mathbf x}
={}&
-\mathbf N^\star\mathbf N^{\star\top}
\left(
\mathbf Q^\star
+\mathbf H^{\star\top}\mathbf K_p\mathbf H^\star
\right)\delta\mathbf x
\\&-\mathbf N^\star\mathbf N^{\star\top}
\mathbf H^{\star\top}\delta\mathbf z
-\mathbf A^{\star\dagger}
\boldsymbol{\Gamma}_{\mathcal A^\star}
\mathbf A^\star\delta\mathbf x,\\
\delta\dot{\mathbf z}
={}&\mathbf K_i\mathbf H^\star\delta\mathbf x.
\end{aligned}
\end{equation}

To separate the normal and tangential directions of the active
inequalities, introduce
\begin{equation}
\label{eq:appendix_coordinates}
\mathbf r:=\mathbf A^\star\delta\mathbf x,
\qquad
\boldsymbol{\psi}:=\mathbf N^{\star\top}\delta\mathbf x.
\end{equation}
The perturbation can then be written as
\(\delta\mathbf x
=\mathbf N^\star\boldsymbol{\psi}
+\mathbf A^{\star\dagger}\mathbf r\). Substituting this expression
into \eqref{eq:appendix_linearized_flow} gives
\begin{subequations}
\label{eq:appendix_decomposed_flow}
\begin{align}
\dot{\mathbf r}
&=-\boldsymbol{\Gamma}_{\mathcal A^\star}\mathbf r,
\label{eq:appendix_r}\\
\dot{\boldsymbol{\psi}}
&=-\mathbf N^{\star\top}
\left(
\mathbf Q^\star
+\mathbf H^{\star\top}\mathbf K_p\mathbf H^\star
\right)
\mathbf N^\star\boldsymbol{\psi}
-\mathbf B^{\star\top}\delta\mathbf z
\nonumber\\
&\quad
-\mathbf N^{\star\top}
\left(
\mathbf Q^\star
+\mathbf H^{\star\top}\mathbf K_p\mathbf H^\star
\right)
\mathbf A^{\star\dagger}\mathbf r,
\label{eq:appendix_xi}\\
\delta\dot{\mathbf z}
&=\mathbf K_i\mathbf B^\star\boldsymbol{\psi}
+\mathbf K_i\mathbf H^\star
\mathbf A^{\star\dagger}\mathbf r.
\label{eq:appendix_z}
\end{align}
\end{subequations}
Since \(\boldsymbol{\Gamma}_{\mathcal A^\star}\succ\mathbf0\), the
inequality-normal subsystem \eqref{eq:appendix_r} is exponentially
stable. It remains to select \(\mathbf K_p\) such that the homogeneous
part of the tangential output--regulation subsystem
\eqref{eq:appendix_xi}--\eqref{eq:appendix_z} is stable.

\medskip

\noindent
\emph{Step 3: Stability of the transverse linearization.}

For every nonzero $\boldsymbol{\psi}$, the vector
$\mathbf d=\mathbf N^\star\boldsymbol{\psi}$ is nonzero and satisfies
$\mathbf A^\star\mathbf d=\mathbf0$. Moreover, since
$\mathbf B^\star=\mathbf H^\star\mathbf N^\star$,
Assumption~\ref{ass:B5} and $k_p>\kappa$ yield
\[
\begin{aligned}
\boldsymbol{\psi}^\top\mathbf R^\star\boldsymbol{\psi}
={}&
\mathbf d^\top
\bigl(
\mathbf Q^\star
+\kappa\mathbf H^{\star\top}\mathbf H^\star
\bigr)\mathbf d
\\
&+(k_p-\kappa)
\|\mathbf B^\star\boldsymbol{\psi}\|^2
>0.
\end{aligned}
\]
Therefore, $\mathbf R^\star\succ\mathbf0$.

The homogeneous part of
\eqref{eq:appendix_xi}--\eqref{eq:appendix_z} has the system matrix
\begin{equation}
\label{eq:appendix_tangent_matrix}
\mathbf J_T
=
\begin{bmatrix}
-\mathbf R^\star&-\mathbf B^{\star\top}\\
\mathbf K_i\mathbf B^\star&\mathbf0
\end{bmatrix}.
\end{equation}
{
Let $s\in\mathbb C$ be an eigenvalue of
\eqref{eq:appendix_tangent_matrix} with eigenvector
\((\boldsymbol{\psi},\boldsymbol{\zeta})\). The corresponding
eigenvalue equations yield
\begin{equation}
\label{eq:appendix_energy_identity}
2\operatorname{Re}(s)
\left(
\|\boldsymbol{\psi}\|^2
+\boldsymbol{\zeta}^*\mathbf K_i^{-1}\boldsymbol{\zeta}
\right)
=-2\boldsymbol{\psi}^*\mathbf R^\star\boldsymbol{\psi}.
\end{equation}
Since \(\mathbf R^\star\succ\mathbf0\),
\eqref{eq:appendix_energy_identity} implies
$\operatorname{Re}(s)\leq0$, with equality only if
$\boldsymbol{\psi}=\mathbf0$. If $s\neq0$, the second
eigenvalue equation associated with
\eqref{eq:appendix_tangent_matrix} then gives
\(\boldsymbol{\zeta}=\mathbf0\), contradicting the definition of an
eigenvector. Hence, every nonzero eigenvalue of \(\mathbf J_T\) has
strictly negative real part.

For $s=0$, \eqref{eq:appendix_energy_identity} gives
\(\boldsymbol{\psi}=\mathbf0\), and
\eqref{eq:appendix_tangent_matrix} then gives
\(\mathbf B^{\star\top}\boldsymbol{\zeta}=\mathbf0\). Consequently,
\begin{equation}
\label{eq:appendix_kernel}
\ker\mathbf J_T
=
\left\{
(\mathbf0,\boldsymbol{\eta}):
\mathbf B^{\star\top}\boldsymbol{\eta}=\mathbf0
\right\}.
\end{equation}
}

The zero eigenvalue in \eqref{eq:appendix_kernel} is semisimple. Indeed,
suppose that \((\mathbf0,\boldsymbol{\eta})\in\ker\mathbf J_T\) also
belongs to \(\operatorname{range}\mathbf J_T\). Then there exist
\(\boldsymbol{\psi}\) and \(\boldsymbol{\zeta}\) such that
\[
\mathbf J_T
\begin{bmatrix}
\boldsymbol{\psi}\\
\boldsymbol{\zeta}
\end{bmatrix}
=
\begin{bmatrix}
\mathbf0\\
\boldsymbol{\eta}
\end{bmatrix}.
\]
The second block equation and
\(\mathbf B^{\star\top}\boldsymbol{\eta}=\mathbf0\) give
$\boldsymbol{\eta}^\top\mathbf K_i^{-1}\boldsymbol{\eta}
=\boldsymbol{\psi}^\top\mathbf B^{\star\top}\boldsymbol{\eta}=0$.
Since \(\mathbf K_i\succ\mathbf0\), this implies
\(\boldsymbol{\eta}=\mathbf0\). Thus,
\(\ker\mathbf J_T\cap\operatorname{range}\mathbf J_T
=\{\mathbf0\}\), which proves that the zero eigenvalue is
semisimple.

Combining \eqref{eq:appendix_r} with the spectral properties of
\eqref{eq:appendix_tangent_matrix}, all nonzero eigenvalues of the
linearized system \eqref{eq:appendix_decomposed_flow} have strictly
negative real parts, and its zero eigenvalue is semisimple. Its zero
eigenspace is
\begin{equation}
\label{eq:appendix_full_kernel}
\left\{
(\mathbf r,\boldsymbol{\psi},\delta\mathbf z):
\mathbf r=\mathbf0,
\ \boldsymbol{\psi}=\mathbf0,
\ \mathbf B^{\star\top}\delta\mathbf z=\mathbf0
\right\}.
\end{equation}

\medskip
\noindent
\emph{Step 4: Local exponential convergence.}

Let
$\mathbf F$ denote the right-hand side of \eqref{eq:proposed_dynamics}.
By Step~1, there exists $\rho>0$ such that $\mathbf F$ is continuously
differentiable on a neighborhood of
$\overline{\mathcal B}_\rho:=\{\boldsymbol\xi:
\|\boldsymbol\xi-\boldsymbol\xi^\star\|\leq\rho\}$. By
Assumption~\ref{ass:B4}, after reducing $\rho$ if necessary,
$\boldsymbol{\mu}_{\mathcal A^\star}(\boldsymbol{\eta})>\mathbf0$ for every
$(\mathbf x^\star,\mathbf z^\star+\boldsymbol\eta)\in\overline{\mathcal B}_\rho$,
and hence
\[
\mathcal E_\rho:=\mathcal E^\star\cap\overline{\mathcal B}_\rho
=\left\{(\mathbf x^\star,\mathbf z^\star+\boldsymbol\eta):
\mathbf B^{\star\top}\boldsymbol\eta=\mathbf0\right\}
\cap\overline{\mathcal B}_\rho .
\]
By Lemmas~\ref{lem:equilibrium_KKT} and~\ref{lem:local_kkt_set}, $\mathcal E_\rho$ is a compact
piece of an affine manifold of equilibria. Its tangent space
$\{(\mathbf0,\boldsymbol\eta):\mathbf B^{\star\top}\boldsymbol\eta=\mathbf0\}$
coincides, under \eqref{eq:appendix_coordinates}, with the zero eigenspace
\eqref{eq:appendix_full_kernel}. Since its dimension equals
$\dim\ker D\mathbf F(\boldsymbol\xi^\star)$, the implicit function theorem
implies that, after reducing $\rho$ if necessary, $\mathcal E_\rho$
contains all equilibria of $\mathbf F$ in $\overline{\mathcal B}_\rho$.

By Step~3 and continuity of $D\mathbf F$, after reducing $\rho$ if
necessary, at every point of $\mathcal E_\rho$ the zero eigenvalue of
$D\mathbf F$ is semisimple with eigenspace equal to the tangent space of
$\mathcal E_\rho$, and all remaining eigenvalues satisfy
$\operatorname{Re}(s)\leq-2\beta$ for some $\beta>0$. The transverse
linearization theorem for manifolds of equilibria therefore yields
$r\in(0,\rho)$ and $c_0>0$ such that every solution with
$\boldsymbol\xi(0)\in\mathcal B_r$ remains in $\overline{\mathcal B}_\rho$
and satisfies
\begin{equation}
\label{eq:appendix_transverse_decay}
\operatorname{dist}\bigl(\boldsymbol\xi(t),\mathcal E_\rho\bigr)
\leq c_0e^{-\beta t}
\operatorname{dist}\bigl(\boldsymbol\xi(0),\mathcal E_\rho\bigr),
\qquad t\geq0.
\end{equation}

We next bound the drift along $\mathcal E_\rho$. Let $L$ be a Lipschitz
constant of $\mathbf F$ on $\overline{\mathcal B}_\rho$. Since $\mathbf F$
vanishes on $\mathcal E_\rho$ and $\overline{\mathcal B}_\rho$ is convex,
$\|\dot{\boldsymbol\xi}(t)\|
\leq L\operatorname{dist}(\boldsymbol\xi(t),\mathcal E_\rho)$. Hence, by
\eqref{eq:appendix_transverse_decay}, for all $0\leq t\leq s$,
\[
\begin{aligned}
\|\boldsymbol\xi(s)-\boldsymbol\xi(t)\|
&\leq\int_t^s Lc_0e^{-\beta\sigma}\,\mathrm d\sigma\,
\operatorname{dist}\bigl(\boldsymbol\xi(0),\mathcal E_\rho\bigr)
\\&\leq\frac{Lc_0}{\beta}e^{-\beta t}
\operatorname{dist}\bigl(\boldsymbol\xi(0),\mathcal E_\rho\bigr).
\end{aligned}
\]
Thus, $\boldsymbol\xi(t)$ is Cauchy and converges to some
$\bar{\boldsymbol\xi}$. By \eqref{eq:appendix_transverse_decay} and
closedness of $\mathcal E_\rho$,
$\bar{\boldsymbol\xi}=(\mathbf x^\star,\bar{\mathbf z})\in\mathcal E_\rho
\subseteq\mathcal E^\star$. Letting $s\to\infty$ gives
\(
\|\boldsymbol\xi(t)-\bar{\boldsymbol\xi}\|
\leq\frac{Lc_0}{\beta}e^{-\beta t}
\operatorname{dist}\bigl(\boldsymbol\xi(0),\mathcal E_\rho\bigr).
\)

Finally, let $\mathcal W:=\mathcal B_{r'}$ with
$r':=\min\{r,\rho/2\}$. For $\boldsymbol\xi(0)\in\mathcal W$, since
$\boldsymbol\xi^\star\in\mathcal E^\star$, every nearest point of
$\mathcal E^\star$ to $\boldsymbol\xi(0)$ lies within distance
$2\|\boldsymbol\xi(0)-\boldsymbol\xi^\star\|<\rho$ of
$\boldsymbol\xi^\star$. Hence,
$\operatorname{dist}(\boldsymbol\xi(0),\mathcal E_\rho)
=\operatorname{dist}(\boldsymbol\xi(0),\mathcal E^\star)$, and
\eqref{eq:local_exp_estimate} follows with $c:=Lc_0/\beta$. Since
$\Sg$ is forward invariant by Proposition~\ref{prop:forward_invariance},
the result holds for all initial states in
$\mathcal W\cap(\Sg\times\R^p)$.

\section{Proof of Theorem~\ref{thm:safe_component_convergence}}
\label{app:proof_safe_convergence}

By Proposition~\ref{prop:forward_invariance} and continuity of the
trajectories, the connected component $\Sg^\star$ is forward invariant.

As in Step~1 of Appendix~\ref{app:proof_local_convergence}, for each
fixed binding set $\mathcal I$, the KKT matrix
\eqref{eq:appendix_qp_kkt_matrix}, with $\mathbf A^\star$ replaced by
$\mathbf G_{\mathcal I}(\mathbf x)$, is nonsingular by
Assumption~\ref{ass:C1}. Hence, with
$\boldsymbol{\lambda}_{\mathcal I^c}^\star=\mathbf0$
by \eqref{eq:cbf_qp_KKTd}, the output--regulation
solution pair
$(\mathbf v^\star,\boldsymbol{\lambda}^\star)$ of the CBF
quadratic program \eqref{eq:cbf_qp} is unique and continuously
differentiable on each region with a fixed binding set. Moreover,
continuity of the quadratic-program KKT condition \eqref{eq:cbf_qp_KKT} ensures
that every one-sided limit is a KKT pair at the transition point,
so the neighboring solutions coincide on their common boundary.
Together with the fact that only finitely many binding sets can occur,
the resulting solution mappings $\mathbf v^\star$ and
$\boldsymbol{\lambda}^\star$ are locally Lipschitz and piecewise
continuously differentiable on
$\Sg^\star\times\R^p$.
Consequently, their compositions with solutions of
\eqref{eq:proposed_dynamics} are locally absolutely continuous and
therefore differentiable almost everywhere.

{
We first clarify that the curvature threshold
\eqref{eq:global_directional_curvature} in Assumption \ref{ass:A8} can be specified before
selecting $k_p$. The mappings $\mathbf v^\star$,
$\boldsymbol\lambda^\star$, and $\mathbf Q$ depend on $k_p$ and
$\mathbf z$ only through the augmented multiplier
$\mathbf z+k_p\mathbf h(\mathbf x)$. For each fixed $\mathbf x$,
this multiplier ranges over $\R^p$ as $\mathbf z$ ranges over $\R^p$,
regardless of $k_p$. Hence, the existence of a uniform $\kappa$
satisfying \eqref{eq:global_directional_curvature} is independent of
the proportional gain used to parameterize these mappings.
We therefore fix any $k_p>\kappa$.
}

Consider
\begin{equation}
\label{eq:global_Vh}
\begin{aligned}
V_h(\mathbf x,\mathbf z)
:=&
\frac{1}{2}\|\mathbf v^\star(\mathbf x,\mathbf z)\|^2
+\frac{1}{2}\mathbf h(\mathbf x)^\top
\mathbf K_i\mathbf h(\mathbf x)
\\&+\sum_{j=1}^m
\lambda_j^\star(\mathbf x,\mathbf z)
\alpha_j\bigl(-g_j(\mathbf x)\bigr).
\end{aligned}
\end{equation}
By \eqref{eq:cbf_qp_KKTc}, $\boldsymbol{\lambda}^\star\geq\mathbf0$.
Hence, since $\mathbf g(\mathbf x)\leq\mathbf0$ on $\Sg^\star$,
one has $V_h\geq0$.

Differentiating \eqref{eq:cbf_qp_KKTa} along
\eqref{eq:proposed_dynamics} and using
\eqref{eq:global_lagrangian_hessian} gives
\(
\dot{\mathbf v}^\star
+\mathbf Q(\mathbf x,\mathbf z)\mathbf v^\star
+\mathbf H(\mathbf x)^\top\mathbf K_p
 \mathbf H(\mathbf x)\mathbf v^\star
+\mathbf H(\mathbf x)^\top\mathbf K_i\mathbf h(\mathbf x)
+\mathbf G(\mathbf x)^\top
 \dot{\boldsymbol{\lambda}}^\star
=\mathbf0.
\)
For each $j$, nonnegativity and local absolute continuity of
$\lambda_j^\star(t)$ imply that $\dot\lambda_j^\star(t)=0$
almost everywhere on $\{t:\lambda_j^\star(t)=0\}$.
Whenever $\lambda_j^\star(t)>0$, quadratic-program complementarity gives
$\nabla g_j(\mathbf x)^\top\mathbf v^\star
=\alpha_j(-g_j(\mathbf x))$. Consequently,
$\dot\lambda_j^\star[\nabla g_j(\mathbf x)^\top\mathbf v^\star
-\alpha_j(-g_j(\mathbf x))]=0$ almost everywhere.
Similarly, quadratic-program complementarity gives
$\lambda_j^\star\nabla g_j(\mathbf x)^\top\mathbf v^\star
=\lambda_j^\star\alpha_j(-g_j(\mathbf x))$.
Using these identities and differentiating \eqref{eq:global_Vh}
along \eqref{eq:proposed_dynamics} yields, almost everywhere,
\begin{equation}
\label{eq:global_Vh_derivative}
\begin{aligned}
\dot V_h
={}&-\mathbf v^{\star\top}
\bigl(\mathbf Q+\kappa\mathbf H^\top\mathbf H\bigr)
\mathbf v^\star
-(k_p-\kappa)\|\mathbf H\mathbf v^\star\|^2\\
&-\sum_{j=1}^m\lambda_j^\star
\alpha_j'(-g_j)\alpha_j(-g_j)\\
\leq{}&-(k_p-\kappa)\|\mathbf H\mathbf v^\star\|^2
-\sum_{j=1}^m\lambda_j^\star
\alpha_j'(-g_j)\alpha_j(-g_j)
\leq0,
\end{aligned}
\end{equation}
where the inequality follows
from  Assumption \ref{ass:A8}, $k_p>\kappa$,
$\boldsymbol\lambda^\star\geq\mathbf0$, and
$\mathbf g(\mathbf x)\leq\mathbf0$.

We next establish boundedness of the trajectory. First, we show that
\begin{equation}
\label{eq:global_equality_error_bound}
\operatorname{dist}(\mathbf x,\mathcal F^\star)
\leq \frac{\|\mathbf h(\mathbf x)\|}{\sigma_h},
\qquad \mathbf x\in\Sg^\star.
\end{equation}
Fix $\mathbf y(0)\in\Sg^\star$ and consider the auxiliary
equality-restoration flow
$\dot{\mathbf y}(s)=-\mathbf H(\mathbf y(s))^\top\mathbf h(\mathbf y(s))$.
Assumption~\ref{ass:global_mfcq} implies the MFCQ for the inequality
constraints alone. Hence, their linearized feasible cone equals the
tangent cone of $\Sg$. By Assumption~\ref{ass:A6},
$\mathbf G_{\mathcal A(\mathbf y)}(\mathbf y)\dot{\mathbf y}\leq\mathbf0$
at every boundary point. The tangency criterion for forward invariance
therefore ensures that $\mathbf y(s)$ remains in $\Sg$. Since
$\mathbf y(s)$ is continuous and starts from $\Sg^\star$, it remains
in the same connected component $\Sg^\star$.

Let $r(s):=\|\mathbf h(\mathbf y(s))\|$. Whenever $r(s)>0$,
Assumption~\ref{ass:C1} gives
\[
\dot{r}(s)
=-\frac{\|\mathbf H(\mathbf y(s))^\top
             \mathbf h(\mathbf y(s))\|^2}{r(s)}
\leq-\sigma_h\|\dot{\mathbf y}(s)\|
\]
and $\dot{r}(s)\leq-\sigma_h^2r(s)$.
If $r(s)=0$, then $\dot{\mathbf y}(s)=\mathbf0$ and the auxiliary solution
remains constant thereafter. Consequently,
\[
\int_0^s\|\dot{\mathbf y}(\tau)\|\,\mathrm d\tau
\leq\frac{r(0)-r(s)}{\sigma_h}
\leq\frac{r(0)}{\sigma_h}.
\]
Thus, the auxiliary solution remains in a bounded ball and cannot have
a finite escape time. It is therefore forward complete.
Moreover, $r(s)\leq r(0)e^{-\sigma_h^2s}$, and the same integral
estimate on $[s_1,s_2]$ shows that $\mathbf y(s)$ is Cauchy as
$s\to\infty$. Its limit belongs to $\mathcal F^\star$, since
$\Sg^\star$ is closed and $\mathbf h(\mathbf y(s))\to\mathbf0$.
Taking $s\to\infty$ in the integral estimate proves
\eqref{eq:global_equality_error_bound}.

Now fix $c\geq0$. Since $\mathbf K_i=k_i\mathbf I$,
\eqref{eq:global_Vh} implies
$\|\mathbf h(\mathbf x)\|\leq\sqrt{2c/k_i}$ whenever
$V_h(\mathbf x,\mathbf z)\leq c$. By
\eqref{eq:global_equality_error_bound} and compactness of
$\mathcal F^\star$, every such point satisfies
\[
\|\mathbf x\|
\leq \max_{\mathbf y\in\mathcal F^\star}\|\mathbf y\|
+\frac{1}{\sigma_h}\sqrt{\frac{2c}{k_i}}.
\]
Thus, the primal projection of each sublevel set of $V_h$ is bounded.

It remains to bound the equality multiplier. Suppose, to the contrary,
that there exists a sequence
$(\mathbf x_k,\mathbf z_k)\in\Sg^\star\times\R^p$ such that
$V_h(\mathbf x_k,\mathbf z_k)$ is bounded while
$\|\mathbf z_k\|\to\infty$. Write
$\boldsymbol{\lambda}_k^\star
=\boldsymbol{\lambda}^\star(\mathbf x_k,\mathbf z_k)$ and
$\rho_k:=\|(\mathbf z_k,\boldsymbol{\lambda}_k^\star)\|$.
The preceding primal bound and closedness of $\Sg^\star$ ensure
that a subsequence satisfies
$\mathbf x_k\to\bar{\mathbf x}\in\Sg^\star$.
Boundedness of $V_h$ in \eqref{eq:global_Vh} gives
$\lambda_{j,k}^\star\alpha_j(-g_j(\mathbf x_k))=O(1)$.
Taking another subsequence,
\[
\frac{1}{\rho_k}
\bigl(\mathbf z_k,\boldsymbol{\lambda}_k^\star\bigr)
\longrightarrow
(\bar{\mathbf z},\bar{\boldsymbol{\lambda}}),
\qquad
\|(\bar{\mathbf z},\bar{\boldsymbol{\lambda}})\|=1,
\]
$\bar\lambda_j=0$ whenever $g_j(\bar{\mathbf x})<0$.
Dividing
\eqref{eq:cbf_qp_KKTa} by $\rho_k$ and passing to the limit yields
\[
\mathbf H(\bar{\mathbf x})^\top\bar{\mathbf z}
+
\mathbf G_{\mathcal A(\bar{\mathbf x})}(\bar{\mathbf x})^\top
\bar{\boldsymbol{\lambda}}_{\mathcal A(\bar{\mathbf x})}
=\mathbf0,
\qquad
\bar{\boldsymbol{\lambda}}_{\mathcal A(\bar{\mathbf x})}\geq\mathbf0.
\]
Taking the inner product with the direction
$\mathbf d_{\bar{\mathbf x}}$ in Assumption \ref{ass:global_mfcq} gives
\(
\bar{\boldsymbol{\lambda}}_{\mathcal A(\bar{\mathbf x})}^\top
\mathbf G_{\mathcal A(\bar{\mathbf x})}(\bar{\mathbf x})
\mathbf d_{\bar{\mathbf x}}
=0.
\)
Because every component of
$\mathbf G_{\mathcal A(\bar{\mathbf x})}
\mathbf d_{\bar{\mathbf x}}$ is strictly negative,
one must have
$\bar{\boldsymbol{\lambda}}_{\mathcal A(\bar{\mathbf x})}=\mathbf0$.
It follows that
$\mathbf H(\bar{\mathbf x})^\top\bar{\mathbf z}=\mathbf0$, and
Assumption~\ref{ass:C1} then gives $\bar{\mathbf z}=\mathbf0$.
Together with $\bar{\boldsymbol{\lambda}}=0$, this contradicts
$\|(\bar{\mathbf z},\bar{\boldsymbol{\lambda}})\|=1$.
Thus, the sublevel sets of $V_h$ on
$\Sg^\star\times\R^p$ are compact. Since $V_h$ is nonincreasing,
every solution remains in its initial compact sublevel set and is
bounded and forward complete.

We now characterize the $\omega$-limit set of the trajectory.
Boundedness and forward completeness imply that this set is nonempty,
compact, and invariant. Since $V_h$ is continuous and nonincreasing,
it is constant on the $\omega$-limit set. Consider any complete
trajectory contained in this set.
Equation~\eqref{eq:global_Vh_derivative},
$k_p>\kappa$, and $\alpha_j'(s)>0$ imply
$\mathbf H(\mathbf x)\mathbf v^\star=\mathbf0$ and
$\lambda_j^\star\alpha_j(-g_j(\mathbf x))=0$ for every $j$.
These identities hold for all times by continuity.

Thus, $\dot{\mathbf h}=\mathbf H(\mathbf x)\mathbf v^\star=\mathbf0$,
so $\mathbf h(\mathbf x(t))\equiv\bar{\mathbf h}$ for some constant
$\bar{\mathbf h}$. Equation~\eqref{eq:proposed_dynamics} then gives
$\mathbf z(t)=\mathbf z(0)+\mathbf K_i\bar{\mathbf h}\,t$.
Since this trajectory lies in a compact set and $\mathbf K_i\succ0$,
one must have $\bar{\mathbf h}=\mathbf0$.
Consequently, \eqref{eq:global_Vh} reduces to
$V_h=\frac12\|\mathbf v^\star\|^2\equiv\bar V_h$ along this
trajectory, where $\bar V_h\geq0$ is constant.

Taking the inner product of \eqref{eq:cbf_qp_KKTa} with
$\mathbf v^\star$ and using
$\mathbf H(\mathbf x)\mathbf v^\star=\mathbf0$ and quadratic-program
complementarity yields
\[
\begin{aligned}
\frac{\mathrm d}{\mathrm dt}f(\mathbf x(t))
&=-\|\mathbf v^\star\|^2
-\boldsymbol\lambda^{\star\top}\mathbf G(\mathbf x)\mathbf v^\star\\
&=-\|\mathbf v^\star\|^2
-\sum_{j=1}^m\lambda_j^\star\alpha_j(-g_j(\mathbf x))
=-2\bar V_h.
\end{aligned}
\]
If $\bar V_h>0$, then
$f(\mathbf x(t))=f(\mathbf x(0))-2\bar V_h t\to-\infty$.
However, $\mathbf x(t)\in\mathcal F^\star$, and continuity of $f$
and compactness of $\mathcal F^\star$ imply that $f$ is bounded below
on this trajectory. Therefore, $\bar V_h=0$, and hence
$\mathbf v^\star=\mathbf0$.

It follows that every point in the $\omega$-limit set satisfies
$\mathbf v^\star=\mathbf0$ and $\mathbf h(\mathbf x)=\mathbf0$,
and is therefore an equilibrium of \eqref{eq:proposed_dynamics}.
By Lemma~\ref{lem:equilibrium_KKT}, its primal component
belongs to $\KKT\cap\Sg^\star$.
Since every bounded trajectory approaches its $\omega$-limit set,
\eqref{eq:global_convergence_conclusion} follows, which completes
the proof.

\section{Proof of Lemma \ref{lem:dynamic_equilibrium_KKT}}
{
Suppose first that
$(\mathbf x^\star,\mathbf z^\star,\boldsymbol{\lambda}^\star)$
is an equilibrium of \eqref{eq:dynamic_icbf_pd}. Since
$\mathbf K_i\succ\mathbf0$, the condition
$\dot{\mathbf z}=\mathbf0$ implies
\(
\mathbf h(\mathbf x^\star)=\mathbf0
\).
Using this equality in $\dot{\mathbf x}=\mathbf0$ gives \eqref{eq:KKTa}.
Moreover, the equilibrium condition for the projected multiplier
dynamics is
\(
\left[
-\boldsymbol{\alpha}
\bigl(-\mathbf g(\mathbf x^\star)\bigr)
\right]^+_{\boldsymbol{\lambda}^\star}
=\mathbf0.
\)
If $\lambda_j^\star>0$, the projection is inactive, and hence
\(
\alpha_{j}\bigl(-g_j(\mathbf x^\star)\bigr)=0.
\) Then
strict monotonicity and $\alpha_j(0)=0$ imply
$g_j(\mathbf x^\star)=0$. If $\lambda_j^\star=0$, the projected
condition gives
\(
-\alpha_{j}\bigl(-g_j(\mathbf x^\star)\bigr)\leq0,
\)
which implies $g_j(\mathbf x^\star)\leq0$. Therefore,
\eqref{eq:KKTb} and \eqref{eq:KKTc} hold
for every $j$, and inequality constraints are satisfied. Recalling \(
\mathbf h(\mathbf x^\star)=\mathbf0
\), 
$(\mathbf x^\star,\mathbf z^\star,\boldsymbol{\lambda}^\star)$
is a KKT triple of \eqref{eq:problem}.

Conversely, suppose that
$(\mathbf x^\star,\mathbf z^\star,\boldsymbol{\lambda}^\star)$
is a KKT triple. Primal feasibility gives
$\mathbf h(\mathbf x^\star)=\mathbf0$, and hence
$\dot{\mathbf z}=\mathbf0$. \eqref{eq:KKTa}, together with $\mathbf h(\mathbf x^\star)=\mathbf0$, yields
$\dot{\mathbf x}=\mathbf0$.
For every $j$ with $\lambda_j^\star>0$, \eqref{eq:KKTc} implies
$g_j(\mathbf x^\star)=0$, and therefore
$\alpha_{j}(-g_j(\mathbf x^\star))=0$. For every $j$ with
$\lambda_j^\star=0$, primal feasibility gives
$g_j(\mathbf x^\star)\leq0$ and hence
$\alpha_{j}(-g_j(\mathbf x^\star))\geq0$. Consequently,
\(
\left[
-\boldsymbol{\alpha}
\bigl(-\mathbf g(\mathbf x^\star)\bigr)
\right]^+_{\boldsymbol{\lambda}^\star}
=\mathbf0,
\)
which gives $\dot{\boldsymbol{\lambda}}=\mathbf0$. Thus,
$(\mathbf x^\star,\mathbf z^\star,\boldsymbol{\lambda}^\star)$
is an equilibrium of \eqref{eq:dynamic_icbf_pd}.
}

\section{Proof of Theorem~\ref{thm:dynamic_icbf_pd_convergence}}
\label{app:proof_DICB_convergence}

By Assumption~\ref{ass:B4}, $\mu_j^\star>0$ for every
$j\in\mathcal A^\star$. For every $j\notin\mathcal A^\star$, one has
$g_j(\mathbf x^\star)<0$, and \eqref{eq:dynamic_icbf_pd_primal} gives
$\dot{\mathbf x}=\mathbf0$ at $\boldsymbol\xi_{\mathrm D}^\star$. Hence,
the $j$th argument of the projected operator in
\eqref{eq:dynamic_icbf_pd_inequality} equals
$-\alpha_j(-g_j(\mathbf x^\star))<0$ at $\boldsymbol\xi_{\mathrm D}^\star$.
By continuity, there exist $\rho>0$ and $c>0$ such that, on
$\overline{\mathcal B}_\rho:=\{\boldsymbol\xi:
\|\boldsymbol\xi-\boldsymbol\xi_{\mathrm D}^\star\|\leq\rho\}$,
$\lambda_j>0$ for every $j\in\mathcal A^\star$ and
$\nabla g_j(\mathbf x)^\top\dot{\mathbf x}-\alpha_j(-g_j(\mathbf x))\leq-c$
for every $j\notin\mathcal A^\star$.

Consequently, while a solution remains in $\overline{\mathcal B}_\rho$,
the projections for $j\in\mathcal A^\star$ are inactive. For
$j\notin\mathcal A^\star$, one has $\tau\dot\lambda_j\leq-c$ whenever
$\lambda_j>0$, and $\dot\lambda_j=0$ whenever $\lambda_j=0$. Thus,
$\boldsymbol\lambda_{(\mathcal A^\star)^c}$ is nonincreasing, reaches zero
no later than
$T_0:=\tau\|\boldsymbol\lambda_{(\mathcal A^\star)^c}(0)\|_\infty/c$, and
remains zero thereafter. On
$\{\boldsymbol\lambda_{(\mathcal A^\star)^c}=\mathbf0\}\cap
\overline{\mathcal B}_\rho$, \eqref{eq:dynamic_icbf_pd} reduces to the
continuously differentiable system in
$(\mathbf x,\mathbf z,\boldsymbol\lambda_{\mathcal A^\star})$ given by
\eqref{eq:dynamic_icbf_pd_primal}, \eqref{eq:dynamic_icbf_pd_equality},
and
$\tau\dot{\boldsymbol\lambda}_{\mathcal A^\star}
=\mathbf G_{\mathcal A^\star}(\mathbf x)\dot{\mathbf x}
-\boldsymbol\alpha_{\mathcal A^\star}
\bigl(-\mathbf g_{\mathcal A^\star}(\mathbf x)\bigr)$,
which we refer to as the reduced system.

Let $\delta\mathbf x$, $\delta\mathbf z$, and
$\delta\boldsymbol{\lambda}_{\mathcal A^\star}$ denote perturbations from
$\boldsymbol\xi_{\mathrm D}^\star$. Linearizing the reduced system at
$\boldsymbol\xi_{\mathrm D}^\star$ gives
\begin{equation}
\label{eq:dynamic_linearized_original}
\begin{aligned}
\delta\dot{\mathbf x}
={}&-
\left(
\mathbf Q^\star
+\mathbf H^{\star\top}\mathbf K_p\mathbf H^\star
\right)\delta\mathbf x
-\mathbf H^{\star\top}\delta\mathbf z
-\mathbf A^{\star\top}
\delta\boldsymbol{\lambda}_{\mathcal A^\star},\\
\delta\dot{\mathbf z}
={}&
\mathbf K_i\mathbf H^\star\delta\mathbf x,\\
\tau\delta\dot{\boldsymbol{\lambda}}_{\mathcal A^\star}
={}&
\mathbf A^\star\delta\dot{\mathbf x}
+
\boldsymbol{\Gamma}_{\mathcal A^\star}
\mathbf A^\star\delta\mathbf x,
\end{aligned}
\end{equation}
where the second term in
the last equation follows from
\(
\delta\!\left[
-\boldsymbol{\alpha}_{\mathcal A^\star}
(-\mathbf g_{\mathcal A^\star})
\right]
=
\boldsymbol{\Gamma}_{\mathcal A^\star}
\mathbf A^\star\delta\mathbf x.
\)

{
Define \(
\delta\boldsymbol{\nu}
:=
\delta\boldsymbol{\lambda}_{\mathcal A^\star}
-\frac1\tau\mathbf A^\star\delta\mathbf x.
\)
Then
\eqref{eq:dynamic_linearized_original} gives
\begin{equation}
\label{eq:dynamic_linearized_transformed}
\begin{aligned}
\delta\dot{\mathbf x}
={}&-
\left(
\mathbf Q^\star
+\mathbf H^{\star\top}\mathbf K_p\mathbf H^\star
+\frac1\tau\mathbf A^{\star\top}\mathbf A^\star
\right)\delta\mathbf x
\\&-\mathbf H^{\star\top}\delta\mathbf z
-\mathbf A^{\star\top}\delta\boldsymbol{\nu},
\\
\delta\dot{\mathbf z}
={}&
\mathbf K_i\mathbf H^\star\delta\mathbf x,\qquad
\delta\dot{\boldsymbol{\nu}}
={}
\frac1\tau
\boldsymbol{\Gamma}_{\mathcal A^\star}
\mathbf A^\star\delta\mathbf x.
\end{aligned}
\end{equation}
}

By Assumption~\ref{ass:B5} and the Finsler lemma, there exists a constant
$\theta>0$ such that
\(
\mathbf Q^\star
+\kappa\mathbf H^{\star\top}\mathbf H^\star
+\theta\mathbf A^{\star\top}\mathbf A^\star
\succ\mathbf0.
\)
Hence, for $k_p>\kappa$ and $0<\tau\leq \tau^\ast:=1/\theta$, we have
\begin{equation}
\label{eq:dynamic_damping_positive}
\mathbf Q^\star
+\mathbf H^{\star\top}\mathbf K_p\mathbf H^\star
+\frac1\tau\mathbf A^{\star\top}\mathbf A^\star
\succ\mathbf0.
\end{equation}
Let $s\in\mathbb C$ be an eigenvalue of the linearized dynamics
\eqref{eq:dynamic_linearized_transformed}, with eigenvector
$(\mathbf u,\mathbf v,\mathbf w)$. The corresponding eigenvalue
equations are
\begin{equation}
\label{eq:eigen_equa}
\begin{aligned}
s\mathbf u
={}&-
\left(
\mathbf Q^\star
+\mathbf H^{\star\top}\mathbf K_p\mathbf H^\star
+\frac1\tau\mathbf A^{\star\top}\mathbf A^\star
\right)\mathbf u
\\&-\mathbf H^{\star\top}\mathbf v
-\mathbf A^{\star\top}\mathbf w,\\
s\mathbf v
={}&\mathbf K_i\mathbf H^\star\mathbf u,\qquad
s\mathbf w
={}\frac1\tau
\boldsymbol{\Gamma}_{\mathcal A^\star}
\mathbf A^\star\mathbf u.
\end{aligned}
\end{equation}
These equations yield
\begin{equation}
\begin{aligned}
&2\operatorname{Re}(s)
\Big(
\|\mathbf u\|^2
+\mathbf v^*\mathbf K_i^{-1}\mathbf v
+\tau\mathbf w^*
\boldsymbol{\Gamma}_{\mathcal A^\star}^{-1}\mathbf w
\Big)\\
={}&-2\mathbf u^*
\Big(
\mathbf Q^\star
+\mathbf H^{\star\top}\mathbf K_p\mathbf H^\star+
\frac1\tau\mathbf A^{\star\top}\mathbf A^\star
\Big)\mathbf u.
\label{eq:dynamic_eigenvalue_identity}
\end{aligned}
\end{equation}
By \eqref{eq:dynamic_damping_positive},
\eqref{eq:dynamic_eigenvalue_identity} implies
$\operatorname{Re}(s)\leq0$, with equality only if
$\mathbf u=\mathbf0$. If $s\neq0$, the remaining two eigenvalue
equations then give $\mathbf v=\mathbf0$ and $\mathbf w=\mathbf0$,
which contradicts the definition of an eigenvector. {Therefore, every
nonzero eigenvalue satisfies $\operatorname{Re}(s)<0$.}

For $s=0$, the zero eigenspace of \eqref{eq:eigen_equa} is
\begin{equation}
\label{eq:dynamic_zero_eigenspace}
\left\{
(\mathbf0,\mathbf v,\mathbf w):
\mathbf H^{\star\top}\mathbf v
+\mathbf A^{\star\top}\mathbf w=\mathbf0
\right\}.
\end{equation}

The zero eigenvalue is semisimple. Indeed, suppose that an element
$(\mathbf0,\mathbf v,\mathbf w)$ of
\eqref{eq:dynamic_zero_eigenspace} also belongs to the range of the
linearized system matrix. Then there exists
$(\widehat{\mathbf u},\widehat{\mathbf v},\widehat{\mathbf w})$
whose image is $(\mathbf0,\mathbf v,\mathbf w)$. By \eqref{eq:eigen_equa},
{
\[
\mathbf v=\mathbf K_i\mathbf H^\star\widehat{\mathbf u},
\qquad
\mathbf w=
\frac1\tau
\boldsymbol{\Gamma}_{\mathcal A^\star}
\mathbf A^\star\widehat{\mathbf u}.
\]
}
Using \eqref{eq:dynamic_zero_eigenspace} gives
{
\begin{align*}
\mathbf v^\top\mathbf K_i^{-1}\mathbf v
+\tau\mathbf w^\top
\boldsymbol{\Gamma}_{\mathcal A^\star}^{-1}\mathbf w
&=
\widehat{\mathbf u}^{\top}
\left(
\mathbf H^{\star\top}\mathbf v
+\mathbf A^{\star\top}\mathbf w
\right)=0.
\end{align*}
}
The positive definiteness of $\mathbf K_i$ and
$\boldsymbol{\Gamma}_{\mathcal A^\star}$ then implies
$\mathbf v=\mathbf0$ and $\mathbf w=\mathbf0$. Hence, the kernel and
range intersect only at the origin, proving that the zero eigenvalue
is semisimple.

By Lemma~\ref{lem:local_kkt_set}, after reducing $\rho$ if necessary,
$\boldsymbol\mu_{\mathcal A^\star}(\boldsymbol\eta)>\mathbf0$ on
$\overline{\mathcal B}_\rho$, and hence
\[
\mathcal E_\rho:=\mathcal E_{\mathrm D}^\star\cap\overline{\mathcal B}_\rho
=\left\{
(\mathbf x^\star,\mathbf z^\star+\boldsymbol\eta,
\boldsymbol\mu(\boldsymbol\eta)):
\mathbf B^{\star\top}\boldsymbol\eta=\mathbf0
\right\}\cap\overline{\mathcal B}_\rho ,
\]
which, by Lemma~\ref{lem:dynamic_equilibrium_KKT}, is a compact
piece of an affine manifold of equilibria of the reduced system. In the
coordinates $(\delta\mathbf x,\delta\mathbf z,
\delta\boldsymbol\lambda_{\mathcal A^\star})$, its tangent space at
$\boldsymbol\xi_{\mathrm D}^\star$ is
\[
\left\{
(\mathbf0,\boldsymbol{\eta},\boldsymbol{\zeta}):
\mathbf H^{\star\top}\boldsymbol{\eta}
+\mathbf A^{\star\top}\boldsymbol{\zeta}=\mathbf0
\right\}.
\]
Since $\delta\mathbf x=\mathbf0$ along this tangent space, we have
$\delta\boldsymbol{\nu}=\delta\boldsymbol{\lambda}_{\mathcal A^\star}$.
Therefore, the tangent space coincides with the zero eigenspace
\eqref{eq:dynamic_zero_eigenspace}, which is semisimple, while all
remaining eigenvalues have negative real parts. Arguing as in Step~4 of
Appendix~\ref{app:proof_local_convergence} for the reduced system, after
reducing $\rho$ if necessary, there exist $r\in(0,\rho)$, $c_1>0$, and
$\beta_\tau>0$ such that every solution of the reduced system starting in
$\mathcal B_r$ remains in $\overline{\mathcal B}_\rho$, converges to some
$\bar{\boldsymbol\xi}\in\mathcal E_\rho$, and satisfies
\begin{equation}
\label{eq:dynamic_reduced_estimate}
\|\boldsymbol\xi(t)-\bar{\boldsymbol\xi}\|
\leq c_1e^{-\beta_\tau t}
\operatorname{dist}\bigl(\boldsymbol\xi(0),\mathcal E_\rho\bigr),
\qquad t\geq0.
\end{equation}

It remains to account for the initial interval on which
$\boldsymbol\lambda_{(\mathcal A^\star)^c}$ may be positive. Let
$d_0:=\operatorname{dist}(\boldsymbol\xi(0),\mathcal E_\rho)$. Since
$\boldsymbol\lambda_{(\mathcal A^\star)^c}=\mathbf0$ on $\mathcal E_\rho$,
one has $\|\boldsymbol\lambda_{(\mathcal A^\star)^c}(0)\|_\infty\leq d_0$,
and hence $T_0\leq\tau d_0/c$. On $[0,T_0]$,
$\boldsymbol\lambda_{(\mathcal A^\star)^c}$ is nonincreasing, while the
remaining components of the vector field of \eqref{eq:dynamic_icbf_pd}
are Lipschitz on $\overline{\mathcal B}_\rho$ and vanish on
$\mathcal E_\rho$. Gronwall's inequality therefore yields a constant
$C_0>0$ such that
$\|\boldsymbol\xi(t)-\boldsymbol\xi(0)\|\leq C_0d_0$ for all
$t\in[0,T_0]$. In particular,
$\boldsymbol\lambda_{(\mathcal A^\star)^c}(T_0)=\mathbf0$ and
$\operatorname{dist}(\boldsymbol\xi(T_0),\mathcal E_\rho)\leq(1+C_0)d_0$.

For $\boldsymbol\xi(0)$ sufficiently close to
$\boldsymbol\xi_{\mathrm D}^\star$, one has
$\boldsymbol\xi(T_0)\in\mathcal B_r$, and the solution coincides with
that of the reduced system for $t\geq T_0$. Applying
\eqref{eq:dynamic_reduced_estimate} from $t=T_0$ gives
\[
\|\boldsymbol\xi(t)-\bar{\boldsymbol\xi}\|
\leq c_1(1+C_0)e^{\beta_\tau T_0}e^{-\beta_\tau t}d_0,
\qquad t\geq T_0,
\]
whereas, for $t\in[0,T_0]$,
\[
\begin{aligned}
\|\boldsymbol\xi(t)-\bar{\boldsymbol\xi}\|
&\leq\|\boldsymbol\xi(t)-\boldsymbol\xi(T_0)\|
+\|\boldsymbol\xi(T_0)-\bar{\boldsymbol\xi}\|
\\&\leq\bigl(2C_0+c_1(1+C_0)\bigr)e^{\beta_\tau T_0}e^{-\beta_\tau t}d_0.
\end{aligned}
\]
Since $T_0\leq\tau\rho/c$, both bounds hold with
$c_\tau:=\bigl(2C_0+c_1(1+C_0)\bigr)e^{\beta_\tau\tau\rho/c}$.
Finally, as in Step~4 of Appendix~\ref{app:proof_local_convergence}, for
$\boldsymbol\xi(0)$ within distance $\rho/2$ of
$\boldsymbol\xi_{\mathrm D}^\star\in\mathcal E_{\mathrm D}^\star$, every
nearest point of $\mathcal E_{\mathrm D}^\star$ to $\boldsymbol\xi(0)$
lies in $\overline{\mathcal B}_\rho$, so
$d_0=\operatorname{dist}(\boldsymbol\xi(0),\mathcal E_{\mathrm D}^\star)$.
Choosing $\mathcal W_\tau$ as a sufficiently small ball around
$\boldsymbol\xi_{\mathrm D}^\star$ proves
\eqref{eq:dynamic_local_exp_estimate}.

\section{Proof of Proposition \ref{prop:dynamic_constraint_violation}}
Fix $\rho>0$ independent of $\tau$ and define
$M:=\|\boldsymbol\mu^\star\|_\infty+\rho$. For each
$\tau\in(0,\tau^\ast]$, let $\mathcal W_\tau$ and $c_\tau$ be given by
Theorem~\ref{thm:dynamic_icbf_pd_convergence}. Since
$\boldsymbol\xi_{\mathrm D}^\star\in\mathcal E_{\mathrm D}^\star$,
\eqref{eq:dynamic_local_exp_estimate} gives, for every solution starting
from $\mathcal W_\tau$,
\[
\|\boldsymbol\xi(t)-\boldsymbol\xi(0)\|
\leq\|\boldsymbol\xi(t)-\bar{\boldsymbol\xi}\|
+\|\bar{\boldsymbol\xi}-\boldsymbol\xi(0)\|
\leq 2c_\tau\|\boldsymbol\xi(0)-\boldsymbol\xi_{\mathrm D}^\star\|.
\]
Define
$\mathcal U_\tau:=\mathcal W_\tau\cap
\{\boldsymbol\xi:\|\boldsymbol\xi-\boldsymbol\xi_{\mathrm D}^\star\|
<\rho/(1+2c_\tau)\}$. Then every solution starting from
$\mathcal U_\tau$ satisfies
$\|\boldsymbol\xi(t)-\boldsymbol\xi_{\mathrm D}^\star\|<\rho$, and hence
$\|\boldsymbol\lambda(t)\|_\infty
\leq\|\boldsymbol\mu^\star\|_\infty+\rho=M$ for all $t\geq0$.

By the definition of the projected operator,
\eqref{eq:dynamic_icbf_pd_inequality} gives, almost everywhere,
\(
\tau\dot\lambda_j
=\bigl[\dot g_j-\alpha_j(-g_j)\bigr]^+_{\lambda_j}
\geq\dot g_j-\alpha_j(-g_j).
\)
Define $\widetilde b_j(t):=-g_j(\mathbf x(t))+\tau\lambda_j(t)$.
Since $\lambda_j(t)\geq0$ and $\alpha_j$ is increasing,
\(
\dot{\widetilde b}_j
\geq-\alpha_j(-g_j(\mathbf x))
\geq-\alpha_j(\widetilde b_j).
\)
The initial conditions imply $\widetilde b_j(0)\geq0$.
Hence, the comparison lemma yields
$\widetilde b_j(t)\geq\eta_j(t;\widetilde b_j(0))\geq0$.
Consequently, $[g_j(\mathbf x(t))]_+\leq\tau\lambda_j(t)$.
Taking the infinity norm 
gives \eqref{eq:dynamic_uniform_violation}.

\bibliographystyle{agsm}
\bibliography{reference}

@article{allibhoy2024control,
  author  = {Ahmed Allibhoy and Jorge Cort{\'e}s},
  title   = {Control-Barrier-Function-Based Design of Gradient Flows for Constrained Nonlinear Programming},
  journal = {IEEE Transactions on Automatic Control},
  volume  = {69},
  number  = {6},
  pages   = {3499--3514},
  year    = {2024},
  doi     = {10.1109/TAC.2023.3306492}
}

@article{Feijer2010Stability,
  author  = {Diego Feijer and Fernando Paganini},
  title   = {Stability of Primal--Dual Gradient Dynamics and Applications to Network Optimization},
  journal = {Automatica},
  volume  = {46},
  number  = {12},
  pages   = {1974--1981},
  year    = {2010},
  doi     = {10.1016/j.automatica.2010.08.011}
}

@article{Cherukuri2017Saddle,
  author  = {Ashish Cherukuri and Bahman Gharesifard and Jorge Cort{\'e}s},
  title   = {Saddle-Point Dynamics: Conditions for Asymptotic Stability of Saddle Points},
  journal = {SIAM Journal on Control and Optimization},
  volume  = {55},
  number  = {1},
  pages   = {486--511},
  year    = {2017},
  doi     = {10.1137/15M1026924}
}

@article{Ames2017Control,
  author  = {Aaron D. Ames and Xiangru Xu and Jessy W. Grizzle and Paulo Tabuada},
  title   = {Control Barrier Function Based Quadratic Programs for Safety Critical Systems},
  journal = {IEEE Transactions on Automatic Control},
  volume  = {62},
  number  = {8},
  pages   = {3861--3876},
  year    = {2017},
  doi     = {10.1109/TAC.2016.2638961}
}

@article{Wabersich2023Predictive,
  author  = {Kim P. Wabersich and Melanie N. Zeilinger},
  title   = {Predictive Control Barrier Functions: Enhanced Safety Mechanisms for Learning-Based Control},
  journal = {IEEE Transactions on Automatic Control},
  volume  = {68},
  number  = {5},
  pages   = {2638--2651},
  year    = {2023},
  doi     = {10.1109/TAC.2022.3175628}
}

@article{Tang2020SemiGlobal,
  author  = {Yujie Tang and Guannan Qu and Na Li},
  title   = {Semi-Global Exponential Stability of Augmented Primal--Dual Gradient Dynamics for Constrained Convex Optimization},
  journal = {Systems \& Control Letters},
  volume  = {144},
  pages   = {104754},
  year    = {2020},
  doi     = {10.1016/j.sysconle.2020.104754}
}

@article{Boob2023Stochastic,
  author  = {Digvijay Boob and Qi Deng and Guanghui Lan},
  title   = {Stochastic First-Order Methods for Convex and Nonconvex Functional Constrained Optimization},
  journal = {Mathematical Programming},
  volume  = {197},
  pages   = {215--279},
  year    = {2023},
  doi     = {10.1007/s10107-021-01742-y}
}

@article{Jia2025FirstOrder,
  author  = {Zhichao Jia and Benjamin Grimmer},
  title   = {First-Order Methods for Nonsmooth Nonconvex Functional Constrained Optimization with or without Slater Points},
  journal = {SIAM Journal on Optimization},
  volume  = {35},
  number  = {2},
  pages   = {1300--1329},
  year    = {2025},
  doi     = {10.1137/23M1569551}
}

@article{Tondel2003Algorithm,
  author  = {Petter T{\o}ndel and Tor Arne Johansen and Alberto Bemporad},
  title   = {An Algorithm for Multi-Parametric Quadratic Programming and Explicit {MPC} Solutions},
  journal = {Automatica},
  volume  = {39},
  number  = {3},
  pages   = {489--497},
  year    = {2003},
  doi     = {10.1016/S0005-1098(02)00250-9}
}

@article{Patrinos2014Accelerated,
  author  = {Panagiotis Patrinos and Alberto Bemporad},
  title   = {An Accelerated Dual Gradient-Projection Algorithm for Embedded Linear Model Predictive Control},
  journal = {IEEE Transactions on Automatic Control},
  volume  = {59},
  number  = {1},
  pages   = {18--33},
  year    = {2014},
  doi     = {10.1109/TAC.2013.2275667}
}

@article{Dhingra2019Proximal,
  author  = {Neil K. Dhingra and Sei Zhen Khong and Mihailo R. Jovanovi{\'c}},
  title   = {The Proximal Augmented Lagrangian Method for Nonsmooth Composite Optimization},
  journal = {IEEE Transactions on Automatic Control},
  volume  = {64},
  number  = {7},
  pages   = {2861--2868},
  year    = {2019},
  doi     = {10.1109/TAC.2018.2867589}
}

@article{Zeng2023Dynamical,
  author  = {Xianlin Zeng and Jinlong Lei and Jie Chen},
  title   = {Dynamical Primal-Dual Nesterov Accelerated Method and Its Application to Network Optimization},
  journal = {IEEE Transactions on Automatic Control},
  volume  = {68},
  number  = {3},
  pages   = {1760--1767},
  year    = {2023},
  doi     = {10.1109/TAC.2022.3152720}
}

@inproceedings{Wang2016MultiObjective,
  author    = {Li Wang and Aaron D. Ames and Magnus Egerstedt},
  title     = {Multi-Objective Compositions for Collision-Free Connectivity Maintenance in Teams of Mobile Robots},
  booktitle = {Proceedings of the 55th IEEE Conference on Decision and Control},
  pages     = {2659--2664},
  year      = {2016},
  doi       = {10.1109/CDC.2016.7798663}
}

@article{Cherukuri2016Asymptotic,
  author  = {Ashish Cherukuri and Bahman Gharesifard and Jorge Cort{\'e}s},
  title   = {Asymptotic Convergence of Constrained Primal--Dual Dynamics},
  journal = {Systems \& Control Letters},
  volume  = {87},
  pages   = {10--15},
  year    = {2016},
  doi     = {10.1016/j.sysconle.2016.01.008}
}

@article{Qu2019Exponential,
  author  = {Guannan Qu and Na Li},
  title   = {On the Exponential Stability of Primal--Dual Gradient Dynamics},
  journal = {IEEE Transactions on Automatic Control},
  volume  = {64},
  number  = {7},
  pages   = {2955--2962},
  year    = {2019},
  doi     = {10.1109/TAC.2018.2863269}
}

@book{Rawlings2017Model,
  author    = {James B. Rawlings and David Q. Mayne},
  title     = {Model Predictive Control: Theory, Computation, and Design},
  publisher = {Nob Hill Publishing},
  year      = {2017}
}

@book{Arrow1958Gradient,
  author    = {Kenneth J. Arrow and Leonid Hurwicz and Hirofumi Uzawa},
  title     = {Studies in Linear and Nonlinear Programming},
  publisher = {Stanford University Press},
  year      = {1958}
}

@article{Niederlander2016Distributed,
  author  = {Simon Niederl{\"a}nder and Jorge Cort{\'e}s},
  title   = {Distributed Coordination for Optimization via Primal--Dual Dynamics},
  journal = {Automatica},
  volume  = {73},
  pages   = {1--9},
  year    = {2016},
  doi     = {10.1016/j.automatica.2016.05.023}
}

@article{Zhang2018Projected,
  author  = {Hong Zhang and Jianwei Wei and Peng Yi and Xiaoqiang Hu},
  title   = {Projected Primal--Dual Gradient Flow of Augmented Lagrangian with Application to Distributed Maximization of the Algebraic Connectivity of a Network},
  journal = {Automatica},
  volume  = {98},
  pages   = {34--41},
  year    = {2018},
  doi     = {10.1016/j.automatica.2018.08.018}
}

@inproceedings{Mestres2023Distributed,
  author    = {Pau Mestres and Jorge Cort{\'e}s},
  title     = {Distributed and Anytime Algorithm for Network Optimization Problems with Separable Structure},
  booktitle = {2023 62nd IEEE Conference on Decision and Control (CDC)},
  year      = {2023}
}

@article{Wang2017Safety,
  author  = {Li Wang and Aaron D. Ames and Magnus Egerstedt},
  title   = {Safety Barrier Certificates for Collision-Free Multirobot Systems},
  journal = {IEEE Transactions on Robotics},
  volume  = {33},
  number  = {3},
  pages   = {661--674},
  year    = {2017},
  doi     = {10.1109/TRO.2017.2665023}
}

@article{Xiao2020HighOrder,
  author  = {Wei Xiao and Calin Belta},
  title   = {High-Order Control Barrier Functions},
  journal = {IEEE Transactions on Automatic Control},
  volume  = {67},
  number  = {7},
  pages   = {3655--3662},
  year    = {2022},
  doi     = {10.1109/TAC.2021.3104938}
}

@article{Blackmore2011Autonomous,
  author  = {Lars Blackmore and Masahiro Ono and Brian C. Williams},
  title   = {Chance-Constrained Optimal Path Planning with Obstacles},
  journal = {IEEE Transactions on Robotics},
  volume  = {27},
  number  = {6},
  pages   = {1080--1094},
  year    = {2011},
  doi     = {10.1109/TRO.2011.2163324}
}

@article{OlfatiSaber2007Consensus,
  author  = {Reza Olfati-Saber and J. Alex Fax and Richard M. Murray},
  title   = {Consensus and Cooperation in Networked Multi-Agent Systems},
  journal = {Proceedings of the IEEE},
  volume  = {95},
  number  = {1},
  pages   = {215--233},
  year    = {2007},
  doi     = {10.1109/JPROC.2006.887293}
}

@book{Bertsekas1999Nonlinear,
  author    = {Dimitri P. Bertsekas},
  title     = {Nonlinear Programming},
  publisher = {Athena Scientific},
  year      = {1999}
}

@article{Nedic2010Distributed,
  author  = {Angelia Nedi{\'c} and Asuman Ozdaglar and Pablo A. Parrilo},
  title   = {Constrained Consensus and Optimization in Networked Systems},
  journal = {IEEE Transactions on Automatic Control},
  volume  = {55},
  number  = {3},
  pages   = {922--938},
  year    = {2010},
  doi     = {10.1109/TAC.2010.2040177}
}

@article{Nedic2018Network,
  author  = {Angelia Nedi{\'c} and Alex Olshevsky and Weijie Shi},
  title   = {Achieving Geometric Convergence for Distributed Optimization over Time-Varying Graphs},
  journal = {SIAM Journal on Optimization},
  volume  = {27},
  number  = {4},
  pages   = {2597--2633},
  year    = {2017},
  doi     = {10.1137/16M1089887}
}

@article{Wabersich2021Learning,
  author  = {Kim P. Wabersich and Melanie N. Zeilinger},
  title   = {A Predictive Safety Filter for Learning-Based Control of Constrained Nonlinear Dynamical Systems},
  journal = {Automatica},
  volume  = {129},
  pages   = {109597},
  year    = {2021},
  doi     = {10.1016/j.automatica.2021.109597}
}

@article{Ma2021SafeMPC,
  author  = {Xin Ma and Yu Song and Francesco Borrelli},
  title   = {A Robust Model Predictive Control Approach for Safety-Critical Systems},
  journal = {Automatica},
  volume  = {123},
  pages   = {109323},
  year    = {2021},
  doi     = {10.1016/j.automatica.2020.109323}
}

@inproceedings{Prajna2007Safety,
  author    = {Stephen Prajna and Ali Jadbabaie},
  title     = {Safety Verification and Reachability Using Barrier Certificates},
  booktitle = {Hybrid Systems: Computation and Control},
  pages     = {477--492},
  year      = {2007},
  doi       = {10.1007/978-3-540-71493-4_37}
}

@article{Boyd2011Distributed,
  author  = {Stephen Boyd and Neal Parikh and Eric Chu and Borja Peleato and Jonathan Eckstein},
  title   = {Distributed Optimization and Statistical Learning via the Alternating Direction Method of Multipliers},
  journal = {Foundations and Trends in Machine Learning},
  volume  = {3},
  number  = {1},
  pages   = {1--122},
  year    = {2011},
  doi     = {10.1561/2200000016}
}

@inproceedings{Ames2019Theory,
  author    = {Aaron D. Ames and Samuel Coogan and Magnus Egerstedt and Gennaro Notomista and Koushil Sreenath and Paulo Tabuada},
  title     = {Control Barrier Functions: Theory and Applications},
  booktitle = {2019 18th European Control Conference (ECC)},
  pages     = {3420--3431},
  year      = {2019}
}

@article{Aubin1991Viability,
  author  = {Jean-Pierre Aubin},
  title   = {Viability Theory},
  journal = {Birkh{\"a}user},
  year    = {1991}
}

@article{Taylor2020Adaptive,
  author  = {Andrew Taylor and Ugo Rosolia and Aaron D. Ames and others},
  title   = {Adaptive Safety-Critical Control with Control Barrier Functions},
  journal = {IEEE Control Systems Letters},
  volume  = {5},
  number  = {2},
  pages   = {561--566},
  year    = {2021}
}

@article{Shi2015Distributed,
  author  = {Wei Shi and Qing Ling and Gang Wu and Wotao Yin},
  title   = {EXTRA: An Exact First-Order Algorithm for Decentralized Consensus Optimization},
  journal = {SIAM Journal on Optimization},
  volume  = {25},
  number  = {2},
  pages   = {944--966},
  year    = {2015}
}

@article{Dawson2023Safe,
  author  = {Christopher Dawson and Sicun Gao and Chuchu Fan},
  title   = {Safe Control with Learned Certificates: A Survey},
  journal = {IEEE Transactions on Robotics},
  year    = {2023}
}

@inproceedings{Havens2021Imitation,
  author    = {Aaron Havens and Bin Hu},
  title     = {On Imitation Learning of Linear Control Policies:
               Enforcing Stability and Robustness Constraints via
               {LMI} Conditions},
  booktitle = {Proceedings of the 2021 American Control Conference},
  pages     = {882--887},
  year      = {2021},
  doi       = {10.23919/ACC50511.2021.9483019}
}

@inproceedings{Tu2022Sample,
  author    = {Stephen Tu and Alexander Robey and Tingnan Zhang
               and Nikolai Matni},
  title     = {On the Sample Complexity of Stability Constrained
               Imitation Learning},
  booktitle = {Proceedings of the 4th Annual Learning for Dynamics
               and Control Conference},
  series    = {Proceedings of Machine Learning Research},
  volume    = {168},
  pages     = {180--191},
  year      = {2022}
}

@article{Rosolia2019Learning,
  author  = {Ugo Rosolia and Francesco Borrelli},
  title   = {Learning Model Predictive Control for Iterative Tasks},
  journal = {IEEE Transactions on Automatic Control},
  volume  = {63},
  number  = {6},
  pages   = {1885--1896},
  year    = {2018}
}

@book{Nagurney1996Projected,
  author    = {Anna Nagurney and Ding Zhang},
  title     = {Projected Dynamical Systems and Variational Inequalities with Applications},
  publisher = {Springer},
  year      = {1996}
}

@article{Allibhoy2025Anytime,
  author  = {Ahmed Allibhoy and Jorge Cort{\'e}s},
  title   = {Anytime Solvers for Variational Inequalities: The (Recursive) Safe Monotone Flows},
  journal = {Automatica},
  volume  = {177},
  pages   = {112287},
  year    = {2025},
  doi     = {10.1016/j.automatica.2025.112287}
}

\end{document}